\documentclass[12pt,preprint]{aastex}

\usepackage{natbib}
\usepackage{ulem}

\usepackage{rotating}
\usepackage{graphicx}

\newcommand{\PSROne}{PSR~J1057$-$5226}
\newcommand{\PSRTwo}{PSR~J1709$-$4429}
\newcommand{\PSRThree}{PSR~J1952$+$3252}
\newcommand{\degree}{\ensuremath{^\circ}}

\slugcomment{Accepted for publication by ApJ }
\begin{document}

\renewcommand{\arraystretch}{0.9}

\title{{\it\bf Fermi} Large Area Telescope Observations of Gamma-ray Pulsars  \\
PSR J1057$-$5226, J1709$-$4429 and J1952$+$3252}
\shorttitle{{\it Fermi} LAT Observations of Three EGRET Pulsars}

\author{
A.~A.~Abdo\altaffilmark{2,3}, 
M.~Ajello\altaffilmark{4}, 
E.~Antolini\altaffilmark{5,6}, 
L.~Baldini\altaffilmark{7}, 
J.~Ballet\altaffilmark{8}, 
G.~Barbiellini\altaffilmark{9,10}, 
M.~G.~Baring\altaffilmark{11}, 
D.~Bastieri\altaffilmark{12,13}, 
K.~Bechtol\altaffilmark{4}, 
R.~Bellazzini\altaffilmark{7}
% ...
}
% ...
\altaffiltext{2}{Space Science Division, Naval Research Laboratory, Washington, DC 20375, USA}
\altaffiltext{3}{National Research Council Research Associate, National Academy of Sciences, Washington, DC 20001, USA}
\altaffiltext{4}{W. W. Hansen Experimental Physics Laboratory, Kavli Institute for Particle Astrophysics and Cosmology, Department of Physics and SLAC National Accelerator Laboratory, Stanford University, Stanford, CA 94305, USA}
\altaffiltext{5}{Istituto Nazionale di Fisica Nucleare, Sezione di Perugia, I-06123 Perugia, Italy}
\altaffiltext{6}{Dipartimento di Fisica, Universit\`a degli Studi di Perugia, I-06123 Perugia, Italy}
\altaffiltext{7}{Istituto Nazionale di Fisica Nucleare, Sezione di Pisa, I-56127 Pisa, Italy}
\altaffiltext{8}{Laboratoire AIM, CEA-IRFU/CNRS/Universit\'e Paris Diderot, Service d'Astrophysique, CEA Saclay, 91191 Gif sur Yvette, France}
\altaffiltext{9}{Istituto Nazionale di Fisica Nucleare, Sezione di Trieste, I-34127 Trieste, Italy}
\altaffiltext{10}{Dipartimento di Fisica, Universit\`a di Trieste, I-34127 Trieste, Italy}
\altaffiltext{11}{Department of Physics and Astronomy, Rice University, MS-108, P. O. Box 1892, Houston, TX 77251, USA}
\altaffiltext{12}{Istituto Nazionale di Fisica Nucleare, Sezione di Padova, I-35131 Padova, Italy}
\altaffiltext{13}{Dipartimento di Fisica ``G. Galilei", Universit\`a di Padova, I-35131 Padova, Italy}
% ...

\begin{abstract}
The {\it Fermi} Large Area Telescope (LAT) data have confirmed the pulsed emission from all six high-confidence gamma-ray pulsars previously known from the EGRET observations. We report results obtained from the analysis of 13 months of LAT data for three of these pulsars (PSR~J1057$-$5226, PSR~J1709$-$4429, and PSR~J1952$+$3252) each  of which had some unique feature among the EGRET pulsars. 
The excellent sensitivity of LAT allows more detailed analysis of the evolution of the pulse profile with energy and also of the variation of the spectral shape with phase. 
We measure the cutoff energy of the pulsed emission from these pulsars for the first time and provide a more complete picture of the emission mechanism. The results confirm some, but not all, of the features seen in the EGRET data.
\end{abstract}



\section{Introduction}

Gamma-ray pulsars offer valuable probes of rotating neutron stars.  The gamma-ray emission is directly related to the primary particle acceleration processes in the pulsar magnetosphere.  This high-energy ($E_\gamma >$100 MeV) radiation represents a significant fraction of the spin-down luminosity of pulsars in some cases.  For a recent summary of implications of gamma-ray pulsar studies, see \citet{Harding2008}.

Until the launch of the  {\it Fermi Gamma-ray Space Telescope} ({\it Fermi}) on 2008 June 11, the most complete results on the highest-energy emission from pulsars came from the Energetic Gamma Ray Experiment Telescope (EGRET) on the Compton Gamma Ray Observatory, which found high-confidence detections of six pulsars \citep{Thompson2008}.  The Large Area Telescope (LAT) on {\it Fermi} has detected a large population of gamma-ray pulsars \citep{LATPULSARS}, including all six of the EGRET pulsars. Detailed studies of the three brightest of these have been reported by the LAT Collaboration: Vela \citep{LATVELA, LATVela2}, Crab \citep{LATCrab}, and Geminga \citep{LATGeminga}. The present paper is a study of the remaining three EGRET high-confidence pulsars, \PSROne\ (B1055$-$52), \PSRTwo\ (B1706$-$44), and \PSRThree\ (B1951+32). The key measured and derived parameters of these pulsars are listed in Table~\ref{Tbl:PSRpars}.


Pulsed emission from \PSROne\ was first detected in radio by~\citet{Vaughan1972} with a period of 197.11~ms. It belongs to the small class of radio pulsars with a strong interpulse midway between the main pulses~\citep{Keith2010}. No evidence of pulsation has been observed in optical, but it has been detected by the \textit{Einstein Observatory}~\citep{Cheng1983} and by {\it EXOSAT}~\citep{Brinkmann1987} as a soft X-ray source. In 1993, pulsed emission  in the X-ray band was detected by {\it ROSAT}~\citep{Oegelman93} and it has been intensively studied with {\it Chandra} ~\citep{Teter2001}. \cite{Becker99} described the unsuccessful search for a pulsar wind nebula around this source.

The geometry of this pulsar has been studied in detail by \citet{Weltevrede2009}, based on Parkes radio observations. Fitting the phase sweep of the polarization position angle to the rotating vector model (RVM)~\citep{radha69}, they found that the pulsar's magnetic axis is inclined at an  angle of $\sim$ 75 $^{\circ}$ to its rotation axis and that the radio interpulse arises from emission formed on open field lines close to the magnetic axis that do not pass through the magnetosphere's null (zero-charge) surface. The radio main pulse emission must originate from field lines lying well outside the polar cap (PC) boundary, beyond the null surface and farther away from the magnetic axis than those of the outer gap (OG) region where the gamma-ray peak is generated.

\PSRTwo\ was first detected as an unidentified gamma-ray source by COS-B~\citep{Swanenburg1981}. Its identification as a pulsar was made about 10 years later with the discovery of pulsed emission from this source in radio wavelengths~\citep{Johnston1992} and with the detection of the gamma-ray pulsations by EGRET using the radio timing information~\citep{Thompson1996}. The pulsed emission in X-rays was found in 2002 with {\it Chandra} ~\citep{Gotthelf2002}. A diffuse pulsar wind nebula was seen around the pulsar in radio~\citep{Frail1994, Giacani2001} and X-rays~\citep{Finley1998, Romani2005} with an extension of $3^{\prime}$ and $110^{\prime\prime}$, respectively. A possible association with a faint supernova remnant (SNR) G342.1$-$2.3, imaged as an arc-like radio structure~\citep{McAdam1993, Frail1994}, was also suggested, although the implication of a high proper motion velocity due to the off-center location of the pulsar in this structure conflicts with the measured scintillation velocity of the pulsar~\citep{Nicastro1996}. No point-like emission was detected in the very-high-energy (VHE) band from the HESS observations at the location of the pulsar~\citep{Aharonian2005-1, Yoshikoshi2009}; however, extended emission was detected from a location offset by $\sim$0.35\degree\ from the pulsar, consistent with the SNR radio location with an extension of 0.3\degree~\citep{Hoppe2009}.


The 39.5 ms radio pulsar \PSRThree\ in the CTB80 SNR was discovered by \cite{Kulkarni1988}. Proper motion measurements of the pulsar using the VLA\footnote{The VLA is operated by the National Radio Astronomy Observatory, which is a facility of the National Science Foundation, operated under cooperative agreement by Associated Universities, Inc.}  \citep{Zeiger2008} leads to a Local Standard Corrected transverse velocity of 274 km s$^{-1}$ and a kinetic age of 51 kyr assuming the pulsar  was born in the geometric center of CTB80 and a distance of 2 kpc. Pulsations in the X-ray domain have been reported from {\it EXOSAT}~\citep{Oegelman1987} and {\it ROSAT}~\citep{Becker1996} with low significance. In the GeV range, a pulsed emission showing two peaks was observed by EGRET \citep{Ramanamurthy1995}. No emission has been detected at higher energy~\citep{Zweerink2009,Albert2007}.


 Each of these pulsars had some unique features among the EGRET pulsars that merit investigation with the greater sensitivity, better resolution, and broader energy range of the LAT.  \PSROne\  in the EGRET data appeared to have the highest efficiency for conversion of spin-down luminosity into gamma-ray energy~\citep{Thompson1999}. \PSRTwo\ had a measured EGRET energy spectrum  that was well described by a broken power law rather than a sharp high-energy cutoff~\citep{Thompson1996}. \PSRThree\ showed an energy spectrum in the EGRET data with no evidence of a cutoff or spectral break out to the 30 GeV limit of the EGRET energy range, although EGRET found only two photons above 10 GeV from this pulsar~\citep{Ramanamurthy1995,Thompson2005}.

In addition to updating conclusions from past observations, the detailed LAT observations of these pulsars have intrinsic interest.  \PSROne\ and \PSRTwo\ are among the minority of gamma-ray pulsars having double peaks in their light curves with a separation less than 0.3 of the phase.  \PSRThree\ is a short-period pulsar and lies in an SNR/pulsar wind nebula complex.

All three of these pulsars are included in the first {\it Fermi} LAT pulsar catalog~\citep{LATPULSARS}.  The present work, in addition to including more data than the catalog paper, also extends the analysis in several ways: (1) a more detailed analysis of the pulse profiles as a function of energy; (2) phase-resolved spectra; (3) a search for off-pulse emission (such as a pulsar wind nebula); and (4) a more detailed discussion of models for beaming and efficiency calculations.

\section{Observations}

\subsection{Gamma-ray Observations}

The LAT aboard {\it Fermi} is an electron-positron pair conversion telescope sensitive to gamma rays with energies in the range from 0.02 to greater than 300 GeV. The LAT is made of a high-resolution silicon micro-strip tracker, a CsI hodoscopic electromagnetic calorimeter, and a segmented plastic scintillator detector to identify the background of charged particles \citep{Atwood2009}. Compared with its predecessor EGRET, the LAT has a larger effective area ($\sim$8000 cm$^{2}$ on-axis, $\geq$1GeV) and improved angular resolution ($\theta_{68} \sim$ 0.6${^\circ}$ at 1 GeV for events in the front section of the tracker). The large field of view ($\sim$2.4 sr) allows the LAT to observe the full sky in survey mode every 3 hr. The LAT timing is derived from a GPS clock on the spacecraft, and gamma rays are hardware time-stamped to an accuracy significantly better than 1 $\mu$s~\citep{LATCalib}. The LAT software tools for pulsars have been shown to be accurate to a few $\mu$s~\citep{smith08}.

\subsection{Radio Observations}

These three pulsars have been monitored by several observatories as part of the pulsar timing campaign for {\it Fermi}, monitoring the $\dot E > 10^{34}$erg s$^{-1}$ pulsars at radio and X-ray wavelengths \citep{smith08}. \PSROne\ and \PSRTwo\ are observed monthly with the 64 m Parkes radio telescope in Australia. Typical observations last for 2 minutes at a frequency of 1.4~GHz (and occasional observations at 0.7 and 3.1~GHz). Full details of the observing and data analysis can be found in \cite{welte2009}. Both pulsars have a high degree of linear polarization in the radio, as is often the case for pulsars with $\dot{E}\gtrsim10^{34}-10^{35}$ erg\,s$^{-1}$~\citep{welte08}. The timing solution for \PSRThree\ uses radio  observations made at the Nan\c{c}ay telescope \citep{Cognard2009} at 1.4~GHz.

\section{Timing Analysis}
\label{Sec:TmgAn}

The radio timing solutions for these three pulsars have been derived from multiple times of arrivals (TOAs) using TEMPO2~\citep{hobbs06}. The TOAs were fit to give each pulsar's spin frequency and frequency derivative. The data are whitened using the fitwaves algorithm within TEMPO2 in order to take into account timing noise. The number of TOAs used in building the timing solutions, the validity period of the solution, and the resulting timing rms for each pulsar are listed in Table~\ref{Tbl:PSRTimeSol}. The complete timing solutions will be made available at the Fermi Science Support Center (FSSC) Web site\footnote{http://fermi.gsfc.nasa.gov/ssc/data/access/lat/ephems/}. When more than one timing solution is available, we have used the one with the lower rms value in order to construct a pulse profile as fine as possible and not to be limited by the accuracy of the timing solution.

In the 13 month interval defined in Section~\ref{Sec:AnalysisDef}, a glitch, a sudden change in the rotation speed of the pulsar, was detected in the analysis of the LAT data for \PSRTwo. While a glitch has been detected from this pulsar previously~\citep{Johnston1995}, this is one of a few detected and characterized by a gamma-ray telescope using gamma-ray data only~\citep{Jackson2002, SazParkinson2009}. The time of the glitch has been narrowed down to a $\sim$22 hr window (MET\footnote{MET is the Mission Elapsed Time in seconds from 00:00 UTC on 2001 January 1.} 240395907-240472803) around 2008 August 14 with the LAT data. The parameters of the glitch were measured from the radio data as $\Delta\nu/\nu=2.7497\pm0.0001\times10^{-6}$ and $\Delta\dot\nu/\dot\nu=4.95\pm0.01\times10^{-3}$~\citep{welte2009}.

\PSRTwo\ was also timed with the LAT data alone obtained over 13 months of LAT observations. The timing solution derived from the LAT observations fits the frequency, frequency derivative, and glitch parameters for this pulsar as well as six WAVE terms to whiten the strong timing noise of this pulsar, using the analysis techniques described in \citet{Ray2010}. We used the radio timing solution with the LAT data to define the radio-gamma offset in the pulse profile constructed with the LAT timing solution.

\section{{\it Fermi} LAT Analysis and Results}
\label{Sec:AnalysisDef}

{\it Fermi} LAT data were analyzed using the standard Science Tools (ST) in its v9r15p2 version available at the FSSC\footnote{http://fermi.gsfc.nasa.gov/ssc/data/analysis/software/}. Only ``Diffuse'' class events, which have the tightest background rejection, were selected for the analysis. In addition, we excluded those photons coming from zenith angles $>105\degree$, where the gamma-rays from Earth's limb produce excessive background contamination. We also have excluded time intervals when the region of interest (ROI) intersects Earth's limb.

The data set spans 13 months from the start of the sky survey observations, 2008 August 4 (MJD 54682.66) to 2009 August 28 (MJD 55071.94). We discarded the observations performed during the  Launch and Early Operations (L\&EO) period in order to avoid the effects of the instrument response function (IRF) changes due to various configuration tests during the L\&EO period~\citep{LATCalib}. The last date of all data sets was determined by the latest available ephemerides provided by the radio telescopes for each of the pulsars reported in this paper. Thus, the data set for each pulsar covers a slightly different time range.

\subsection{Pulse Profiles}
\label{Sec:PlsProfile}

Like all pair-production detectors, the LAT's angular resolution is dominated by multiple scattering at low photon energies. It is therefore important to use an energy-dependent radius similar to the LAT point spread function (PSF) in order to include only the events that can be strongly associated with that source. The accurate parameterization of the LAT PSF to be used for science analysis is described by the IRFs. A simplified, acceptance-averaged approximation of the PSF with  a $68\%$ containment angle is given by $<\theta_{68}(E)> = (0.8^\circ) \times (E_{{\rm GeV}})^{-0.8}$. Accordingly, for the pulse profile analysis, we selected only those events within an energy-dependent radius of $\theta < {\rm Max}( {\rm Min}(R_{\rm max}, \theta_{68}), 0.35\degree)$ around each pulsar: the minimum value of 0.35\degree\ was set in order to be sure to keep all high-energy photons~\citep{LATPULSARS}, while a maximum radius, $R_{\rm max}$, was introduced to reduce the background contamination at low energies. The value of $R_{\rm max}$ was determined such that it maximizes the $H$-test~\citep{deJager1989} performed on the light curve built with only low-energy photons ($0.1\,{\rm GeV}< E <0.3\,{\rm GeV}$): this analysis found $R_{\rm max}$ of 2.0\degree\ for \PSROne\ and \PSRThree\, and 2.4\degree\ for \PSRTwo. Note that this ROI cut was only used to select data for obtaining the pulse profiles, but was not applied to the data sample used in the spectral fits.

Finally, photons with energies above 100 MeV were used for this analysis. The energy cut is necessary because all three pulsars are close to the Galactic plane, and due to the poor angular resolution below 100\,MeV the photons from the strong Galactic diffuse emission hide any periodicity.

In the following subsections, we report the pulse profile characteristics for each pulsar. We give the total number of pulsed and background photons obtained with the selection above. For each pulsar, we show a full-band pulse profile and its evolution with energy, constructing pulse profiles for four different energy bands of 0.1$-$0.3\,GeV, 0.3$-$1.0\,GeV, 1.0$-$3.0\,GeV, and $>$3.0\,GeV  for \PSROne\ and \PSRThree; and for five energy bands of 0.1$-$0.3\,GeV, 0.3$-$1.0\,GeV, 1.0$-$3.0\,GeV, 3.0$-$10.0\,GeV, and $>$10.0\,GeV for \PSRTwo. These pulse profiles have been built with a fixed bin width of 0.01 in phase except for the highest energy band for which we have used a bin width of 0.02 because of limited statistics. In these gamma-ray profiles, $\varphi=0$ is defined as the maximum of the main radio peak for \PSRTwo\ and \PSRThree. The radio light curve for \PSROne\ exhibits multiple peaks and we have chosen to put the radio interpulse peak at phase zero. We hence shifted the gamma-ray profile accordingly. The pulse shapes in each band are parameterized by fitting with simple functions, and the evolution of these parameters with energy is discussed for each pulsar.

\subsubsection{PSR~J1057$-$5226}

Using the data selection and energy-dependent ROI described in Sections~\ref{Sec:AnalysisDef} and \ref{Sec:PlsProfile}, we detected 6106 events with energies above 100 MeV in the region around the pulsar. Since there is no evidence of significant off-pulse emission (see Section~\ref{Sec:OffEm}), we evaluated the number of background photons from the off-pulse phase region, and found $2834 \pm 80$ background photons and $3272 \pm 112$ pulsed photons.

\begin{figure}[phtb]
  \centering
  \includegraphics[width=1.0\textwidth]{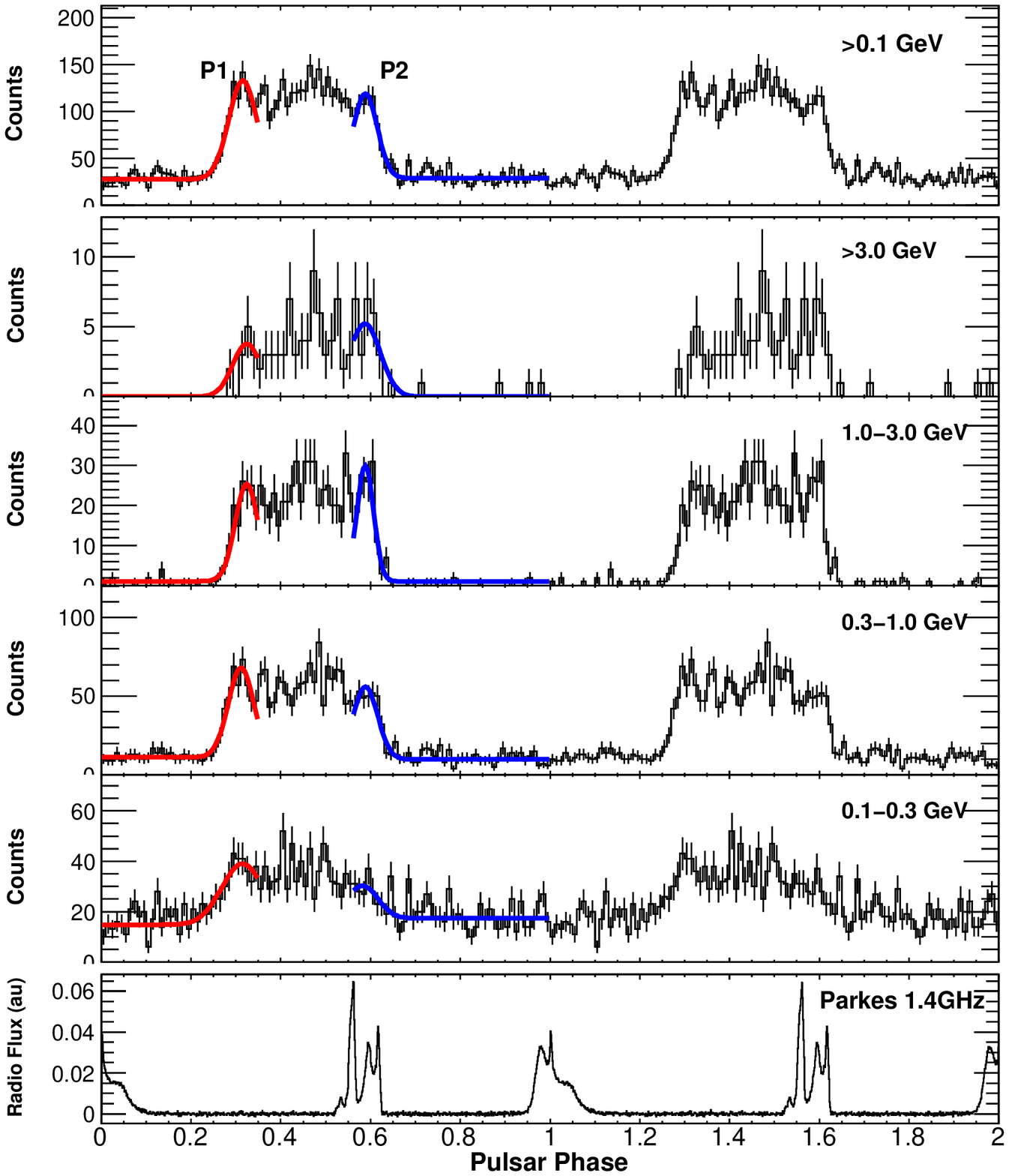}
  \caption{Energy dependence of the \PSROne\ pulse profile. The top panel is the pulse profile in the full energy band; the following four panels show the pulse profile in five different energy bands: $0.1-0.3$\,GeV, $0.3-1.0$\,GeV, $1.0-3.0$\,GeV, and $>3.0$\,GeV. The bin widths are 0.01 in phase except the highest energy band which is 0.02. Two rotation cycles are shown, and the fitting functions (thick solid lines) are superimposed on the light curves in the first cycle. The bottom panel shows the pulse profile at radio wavelengths, in arbitrary units (au), at 1.4~GHz provided by Parkes radio telescope.}
  \label{Fig:J1057PlsProfEDept}
\end{figure}

\begin{figure}[phtb]
  \centering
  \includegraphics[width=0.85\textwidth]{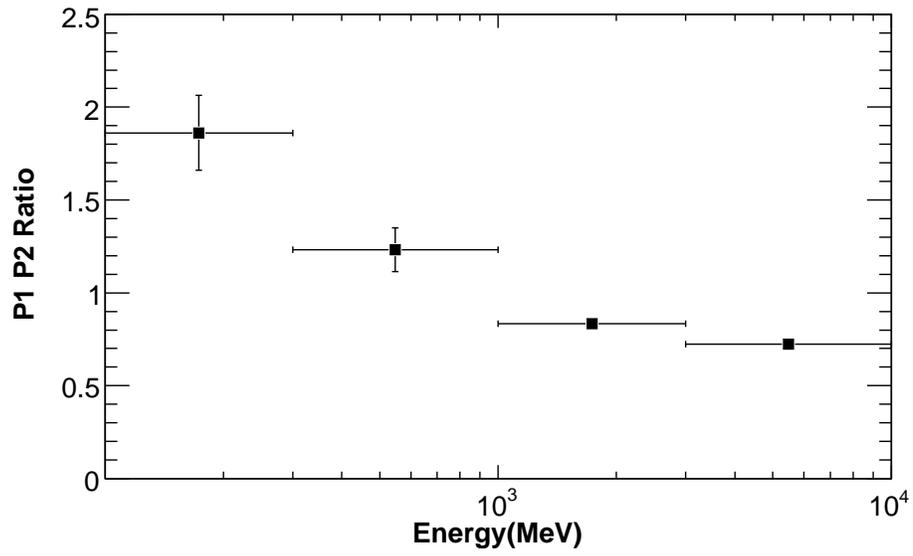}
\caption{Evolution of leading peak to trailing peak ratio with energy for \PSROne.}
  \label{Fig:J1057PlsProfEDeptTrend}
\end{figure}

The pulse profiles in different energy bands are shown in Figure~\ref{Fig:J1057PlsProfEDept} for two complete pulsar rotation cycles. The complex pulsed emission profile extends from 0.25 to 0.65 and changes significantly with energy.  Although they do not represent a major portion of the emission, small leading (P1) and trailing (P2) peaks are always visible. We fitted only these two peaks with Gaussians over a constant background. The position of P1 is $0.31 \pm 0.01$ and its half-width is $0.04 \pm 0.01$ while the position of P2 is $0.59 \pm 0.01$ with a half width of $0.03 \pm 0.01$. The location and half-widths of P1 and P2 obtained from the fits in all energy bands are given in Table~\ref{Tbl:J1057PlsFit}. These two peaks are separated by $0.28 \pm 0.03$ in phase. The region between these two peaks is complex. More photons will be needed to resolve the shape. The positions of the two peaks are constant with energy within the fit uncertainties. Figure~\ref{Fig:J1057PlsProfEDeptTrend}  shows the P1/P2 ratio, seen to decrease with increasing energy. From Figure~\ref{Fig:J1057PlsProfEDept}, P2 is $0.03 \pm 0.01$ after the peak of the radio main pulse and P1 is $0.75 \pm 0.03$ after it. We find 333 photons with energies greater than 3\,GeV, and the highest energy photon consistent with the pulsar position comes from the trailing peak, with a phase of 0.61 and an energy of 8.7\,GeV.

\subsubsection{PSR~J1709$-$4429}
\label{Sec:GammaObsPSR1709}

Due to timing uncertainty around the glitch,  we have chosen to exclude all the photons detected before the end of the glitch relaxation time on 2008 September 16 (MET 243216004). Using the data selection and energy-dependent ROI described previously, we detected 30448 events with energies above 100 MeV in the region around \PSRTwo.  Failing to detect any significant off-pulse source in the phase region $\varphi= 0.7-1.0$ (Section~\ref{Sec:OffEm}), we estimated the number of photons due to the background as $ 17004 \pm 130$ from this off-pulse phase range.

\begin{figure}[ptbh]
 \centering
 \includegraphics[width=1.0\textwidth]{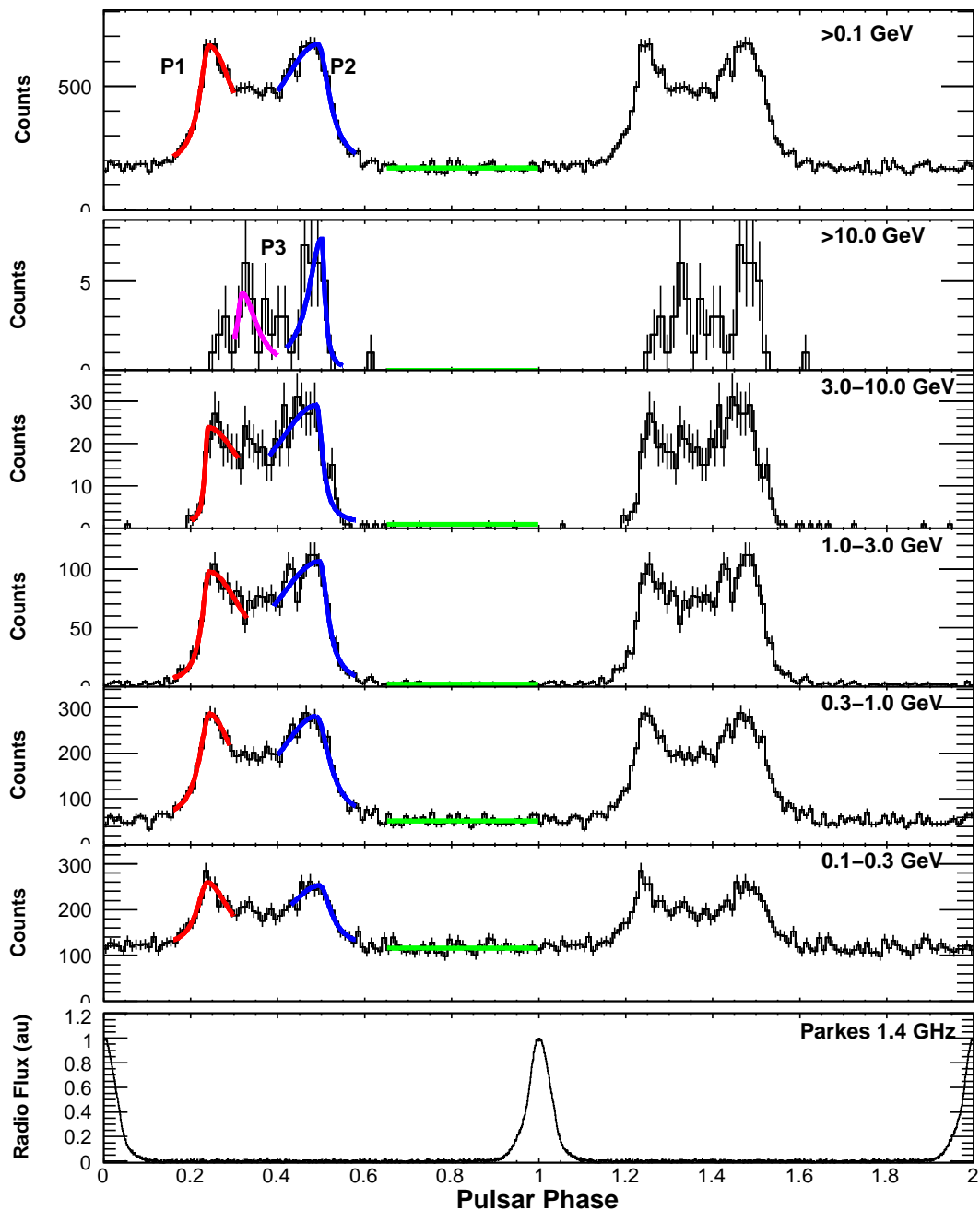}
 \caption{Energy dependence of the \PSRTwo\ pulse profile. The top panel is the pulse profile in the full energy band; the following five panels show the pulse profile in five different energy bands as indicated. The bin widths are 0.01 in phase except the highest energy band which is twice this width. Two rotation cycles are shown, and the fitting functions (thick solid lines) are superimposed on the light curves in the first cycle. The bottom panel shows the pulse profile at radio wavelengths at 1.4~GHz provided by the Parkes radio telescope.}
 \label{Fig:J1709PlsProfEDept}
\end{figure}

Figure~\ref{Fig:J1709PlsProfEDept} shows the pulse profiles in separate energy bands for two complete pulsar rotation cycles. The full-energy-band pulse profile shows a dominant two-peak structure as opposed to the EGRET pulse profile which could not differentiate between two or three peaks. The first peak (P1) is located at $0.242 \pm 0.002$ in phase and the second peak (P2) is located at $0.492 \pm 0.004$ in phase. Both peaks are strongly asymmetric with sharp outer edges toward the off-pulse region with wider and structured inner edges toward the bridge region; hence, both peaks were characterized with asymmetric Lorentzian functions, which have different widths for the leading and trailing edges, as described in \citet{LATVela2}. The Lorentzian half-widths for the outer edges are $0.027 \pm 0.002$ and $0.032 \pm 0.003$, and the Lorentzian half-widths for the inner edges are $0.072 \pm 0.008$  and $0.118 \pm 0.013$ for P1 and P2, respectively. The location and half-widths of P1 and P2 obtained from the fits in all energy bands are given in Table~\ref{Tbl:J1709PlsFit}.

\begin{figure}[phtb]
  \centering
  \includegraphics[width=0.85\textwidth]{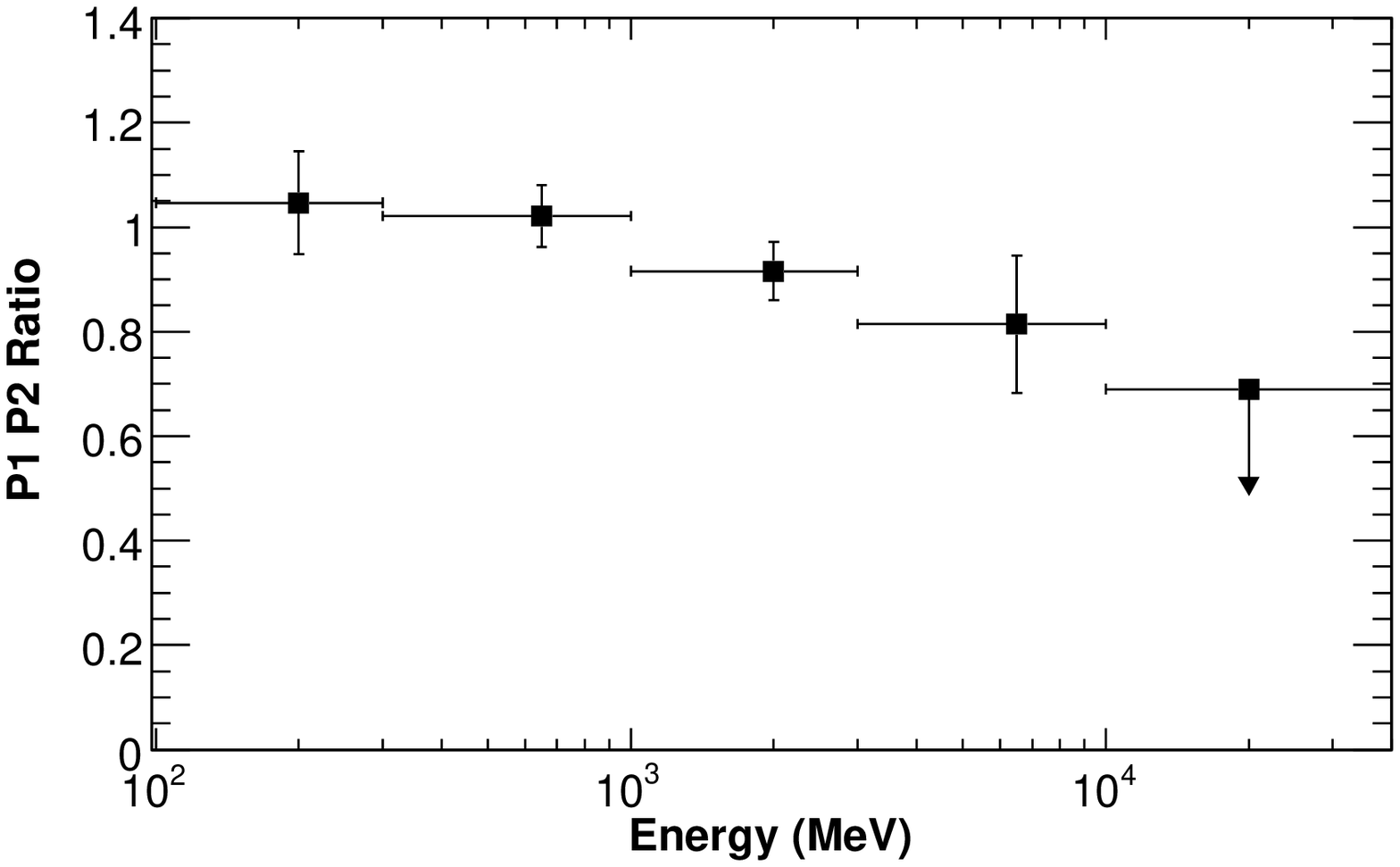}
  \caption{Evolution of the P1 to P2 ratio with energy for \PSRTwo.}
  \label{Fig:J1709P1P2}
\end{figure}

The locations of the first and second peaks (P1 and P2) stay constant over the energy bands, but P1 seems to disappear in the highest energy band, $>10$ GeV. The ratio of the first peak to the second peak slowly decreases with energy, as seen in Figure~\ref{Fig:J1709P1P2}. The pulse profile shows additional peak-like features between the peaks in the three highest energy bands. A third peak (P3) in the bridge region at the phase of 0.35 is seen above 3 GeV, similar to what is observed in the Vela pulsar, and P3 becomes as large as P2 above 10 GeV. However, no phase shift of P3 with energy is detected for this pulsar with the current statistics.  Above 10 GeV, 45 photons were detected from the direction of the pulsar in the pulsed region, $\varphi= 0.1-0.7$. The highest energy photon has an energy of 47.7 GeV and is detected at phase 0.31.  The approximate radio-P1 lag was found to be 0.24 using the method discussed in Section~\ref{Sec:TmgAn}.

\subsubsection{PSR~J1952$+$3252}

Using the data selection and energy-dependent ROI described in Sections~\ref{Sec:AnalysisDef} and \ref{Sec:PlsProfile}, we detected 10314 photons with energies above 100 MeV in the region around this pulsar.  The number of background photons, estimated assuming no emission from the off-pulse region, $\varphi= 0.8-1.0$, as discussed in Section~\ref{Sec:OffEm}, is 8308 $\pm$ 159.

\begin{figure}[!ph]
  \centering
  \includegraphics[scale=0.8]{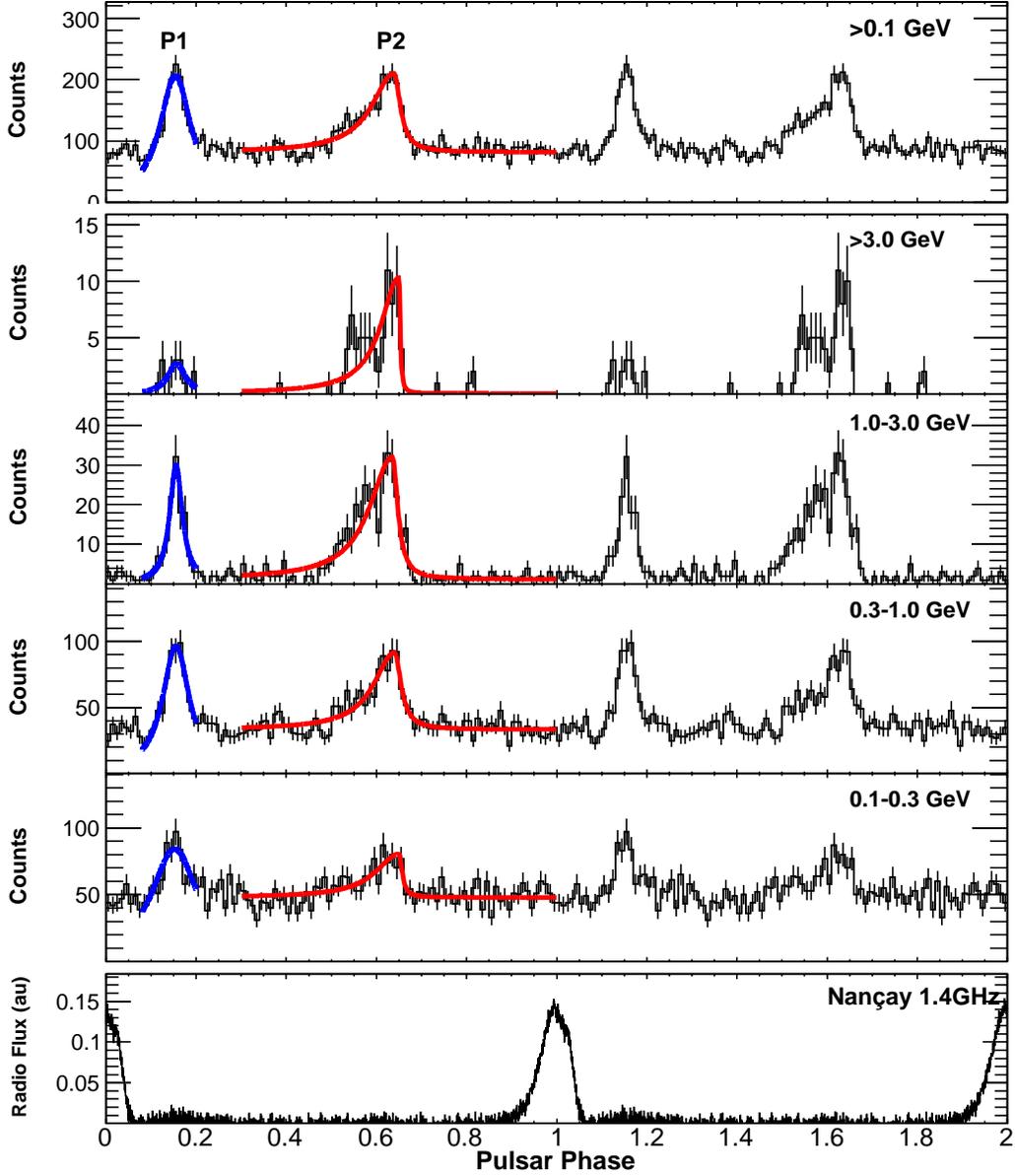}
  \caption{\PSRThree\ light curves. The top panel is the gamma-ray pulse profile for $E>$0.1 GeV within the energy-dependent 68\% containment radius around the pulsar position. Each bin is 0.01 in phase. Two rotation cycles are shown and the fitting functions (thick solid lines) are superimposed on the light curves in the first cycle. Four following frames: light curves in the indicated energy ranges.  The bottom panel is the radio profile at 1.4 GHz obtained at the Nan\c cay radio telescope.}
  \label{lightcurves1952}
\end{figure}

\begin{figure}[ph]
 \centering
  \includegraphics[width=0.85\textwidth]{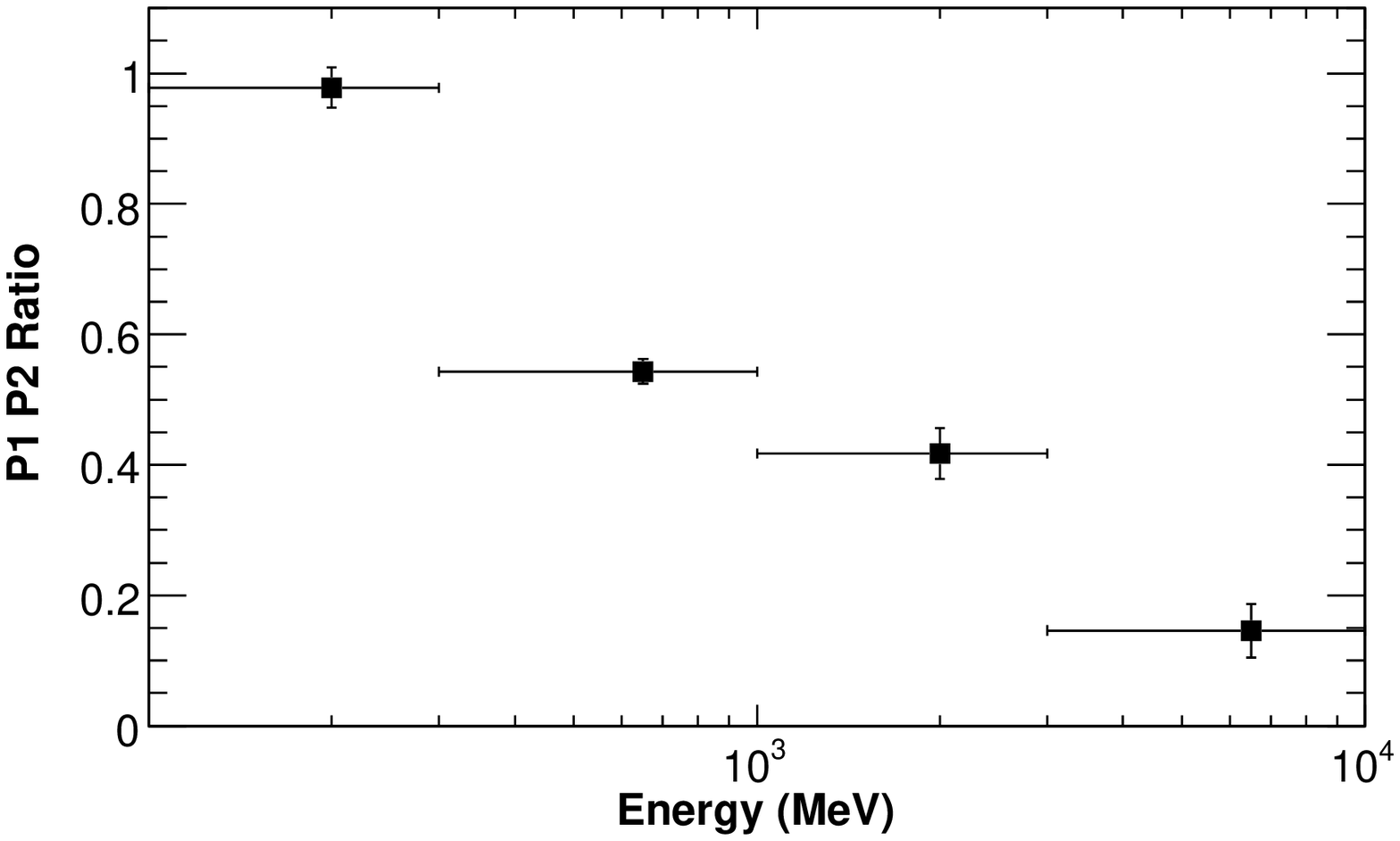}
\caption{Evolution of the P1 to P2 ratio with energy for \PSRThree.}
 \label{fit_results_1952}
\end{figure}

Figure~\ref{lightcurves1952} shows the energy evolution of the pulse profile, which exhibits two clear peaks (further referred to as P1 and P2).  P1 appears symmetric over all energy ranges whereas P2 is clearly asymmetric; accordingly P1 is fitted with a symmetric Lorentzian and P2 with an asymmetric Lorentzian function over a constant background.  The fitted functions are shown in Figure~\ref{lightcurves1952}, and the location and half-widths of P1 and P2 obtained from these fits in all energy bands are given in Table~\ref{Tbl:J1952PlsFit}. The peak positions are independent of the photon energies and the FWHMs of the peaks decrease with increasing energy.  Figure~\ref{fit_results_1952} plots the evolution of the P1/P2 ratio through the four energy bands defined previously, showing that P1 fades with increasing energy whereas P2 persists. The highest energy photon in the whole sample is 25.7 GeV and belongs to P2 at $\varphi = 0.58$, and nine events with $E > 10$\,GeV have been detected.

P1 lags the maximum of the 1.4 GHz radio peak shown in the bottom frame of Figure~\ref{lightcurves1952} by 0.154 $\pm$ 0.001. The separation of the gamma-ray peaks is 0.485 $\pm$ 0.017.

\subsection{Spectral Analysis}
\label{Sec:SpectAn}

For the spectral analysis of each pulsar, we used a data set that covers the same time period as the pulse profile studies and included all photons with energies in the range $100\,\rm{MeV}-100\,\rm{GeV}$ in the 10\degree\ region around each pulsar position. The ``P6\_V3'' IRFs, which are a post-launch update to address gamma-ray detection inefficiencies correlated with the trigger rate, were used in the analysis. The systematic errors on the effective area are $\le$ 5\% near 1 GeV, 10\% below 0.1 GeV, and 20\% over 10 GeV. In order to propagate the uncertainties on the effective area to the systematic errors on the three spectral parameters, the index, flux, and the energy cutoff, six ``modified IRF'' sets were used, each pair bracketing the nominal (P6\_V3) effective area by a correction factor derived corresponding to each spectral parameter.

\begin{figure}[ph]
\centering
  \begin{tabular}{c}

    \includegraphics[width=0.85\textwidth]{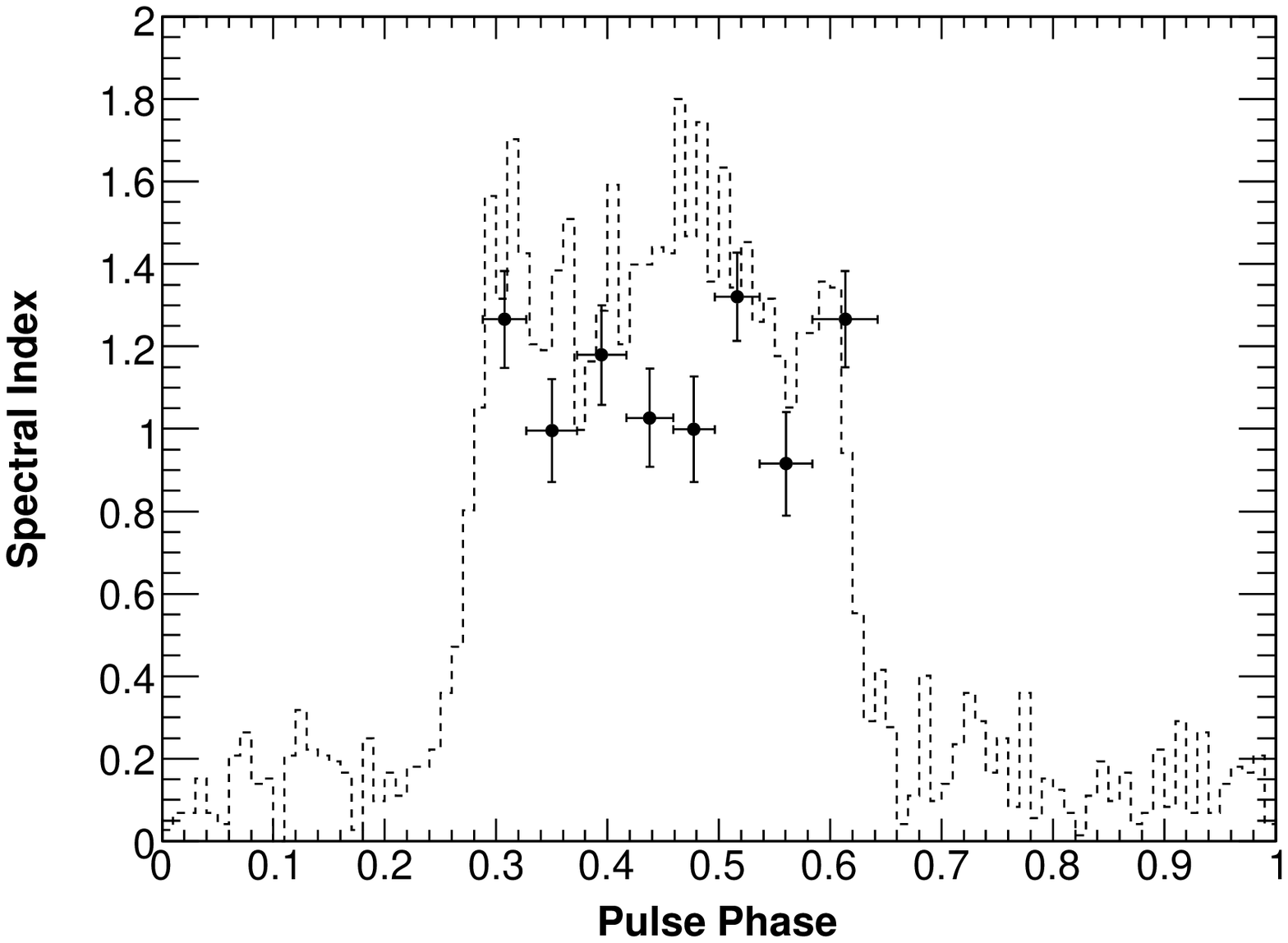}\\
    \includegraphics[width=0.85\textwidth]{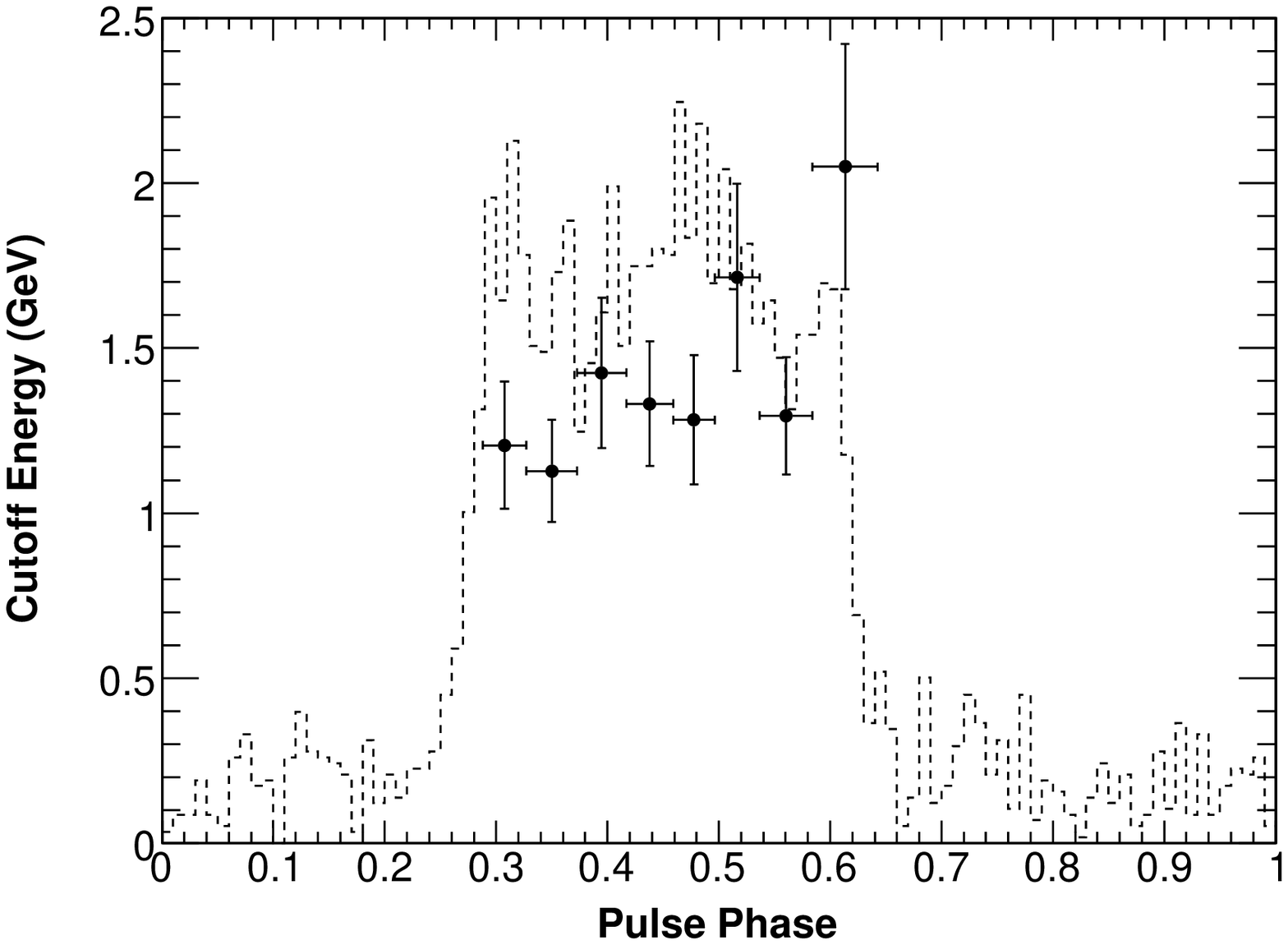}
  \end{tabular}
  \caption{Evolution of the photon index (top panel) and the cutoff energy (bottom panel) of the pulsed emission spectrum through the pulse profile of the \PSROne.}
  \label{B1055_PhaseScan_Plots}
\end{figure}

\begin{figure}[ph]
 \centering
 \includegraphics[width=0.85\textwidth]{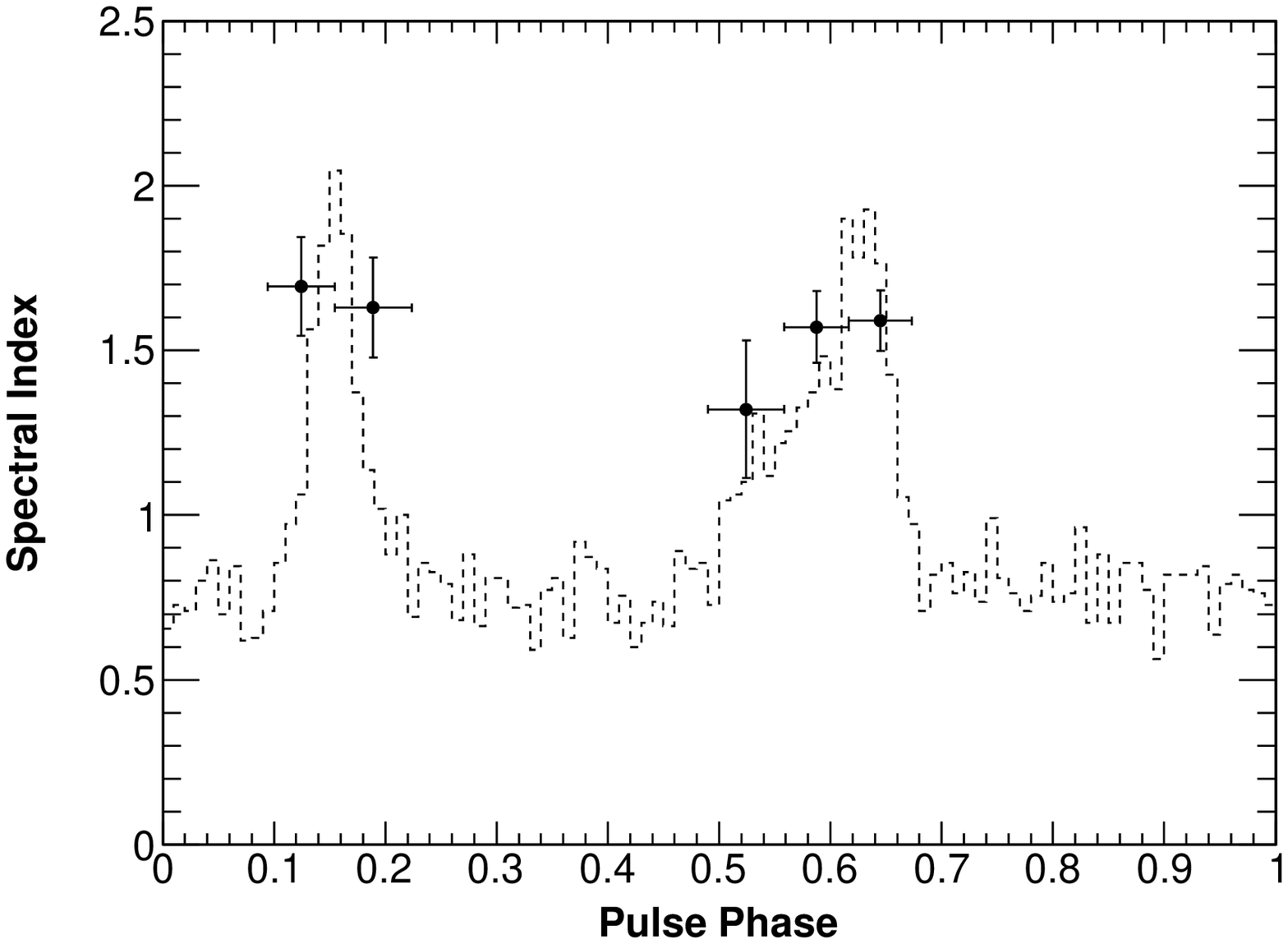}
 \includegraphics[width=0.85\textwidth]{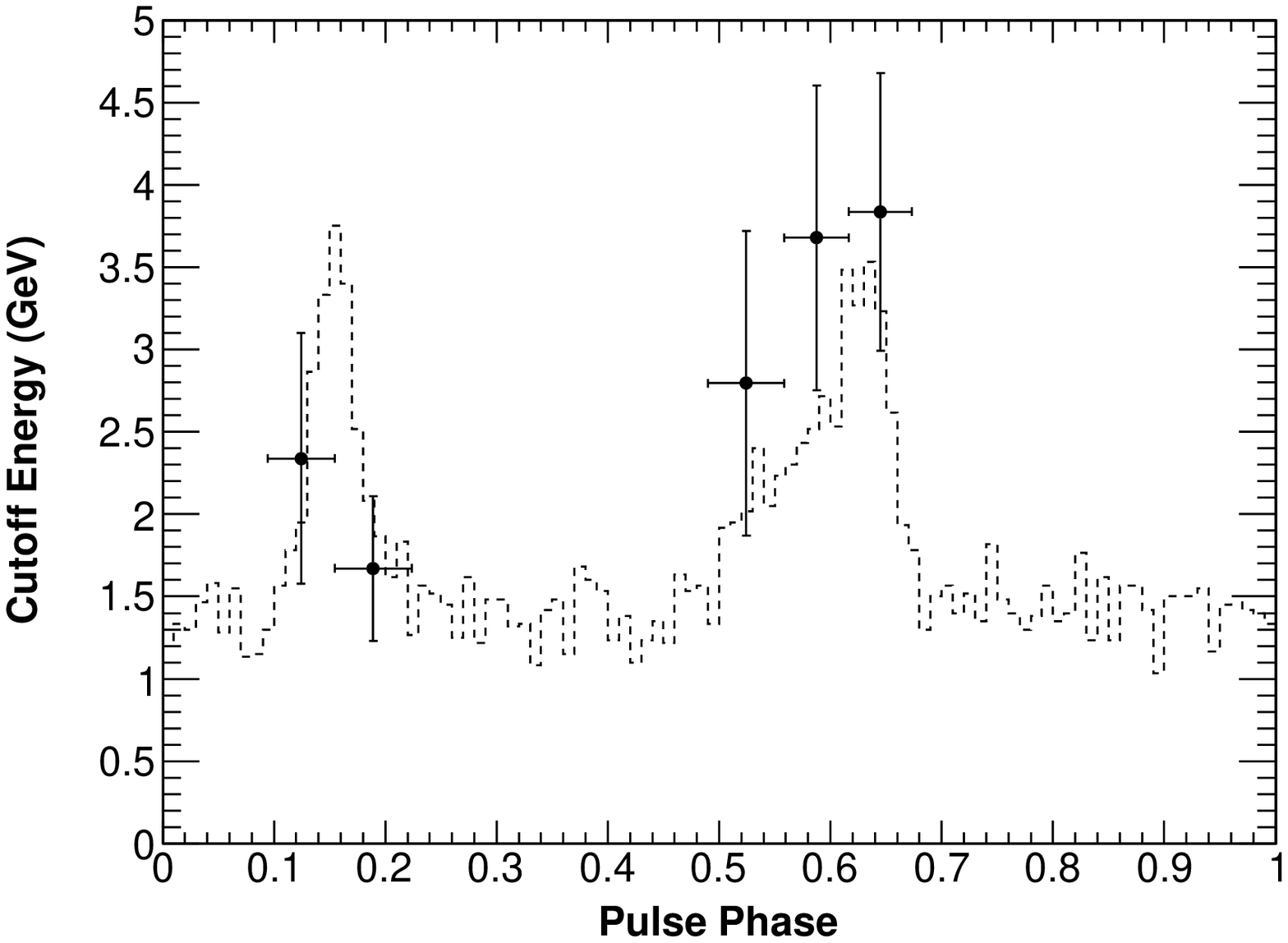}
 \caption{\PSRThree: evolution of the photon index (top panel) and the cutoff energy (bottom) with phase. Only statistical errors are shown.}
 \label{phaseresolved_j1952}
\end{figure}

We measured a phase-averaged spectrum for each pulsar, and the results are reported in Section~\ref{Sec:PASpect}. All three pulsars were sufficiently bright that it was also possible to measure phase-resolved spectra for them. The phase-resolved spectra for these pulsars are reported in Section~\ref{Sec:PRSpect}. Phase-averaged and phase-resolved spectra for the three pulsars are shown in Figures~\ref{sed1055_acdc} and \ref{B1055_PhaseScan_Plots} for \PSROne, in Figures~\ref{SedJ1709} and \ref{Fig:J1709PhRslvd} for \PSRTwo, and in Figures~\ref{sed1952} and \ref{phaseresolved_j1952} for \PSRThree.  

\subsubsection{Phase-averaged Spectrum}
\label{Sec:PASpect}
The emission spectrum for these pulsars in the energy range from 100 MeV to 100 GeV was derived using the LAT tool {\it gtlike}, part of the ST package, based on a maximum-likelihood method~\citep{mattox96}. This tool fits a model representing the point sources and diffuse emission in the selected ROI to the data and finds the best-fit parameters to optimize the likelihood function describing the data. The Galactic diffuse emission is modeled by the standard mapcube file gll\_iem\_v2.fits. A tabulated model for the isotropic component, isotropic\_iem\_v02.txt, representing the extragalactic emission, as well as the residual instrumental background, was used. Both diffuse models are publicly released by the FSSC. The absolute normalization was left as a free parameter for both models during the likelihood fit. The nearby sources were taken from the First LAT Catalog~\citep{1FGLCat}. We included all sources within 17\degree\ of the pulsars. The spectra of all the nearby sources except pulsars were modeled with a power-law spectral shape.  Pulsars were modeled as power laws with exponential cutoffs.  The spectral parameters for sources within 10\degree\ of the pulsar were left free in the fit, and the spectral parameters for other sources were fixed to the spectral parameters obtained from the catalog run.

\begin{figure}[ph]
 \centering
 \includegraphics[scale=0.8]{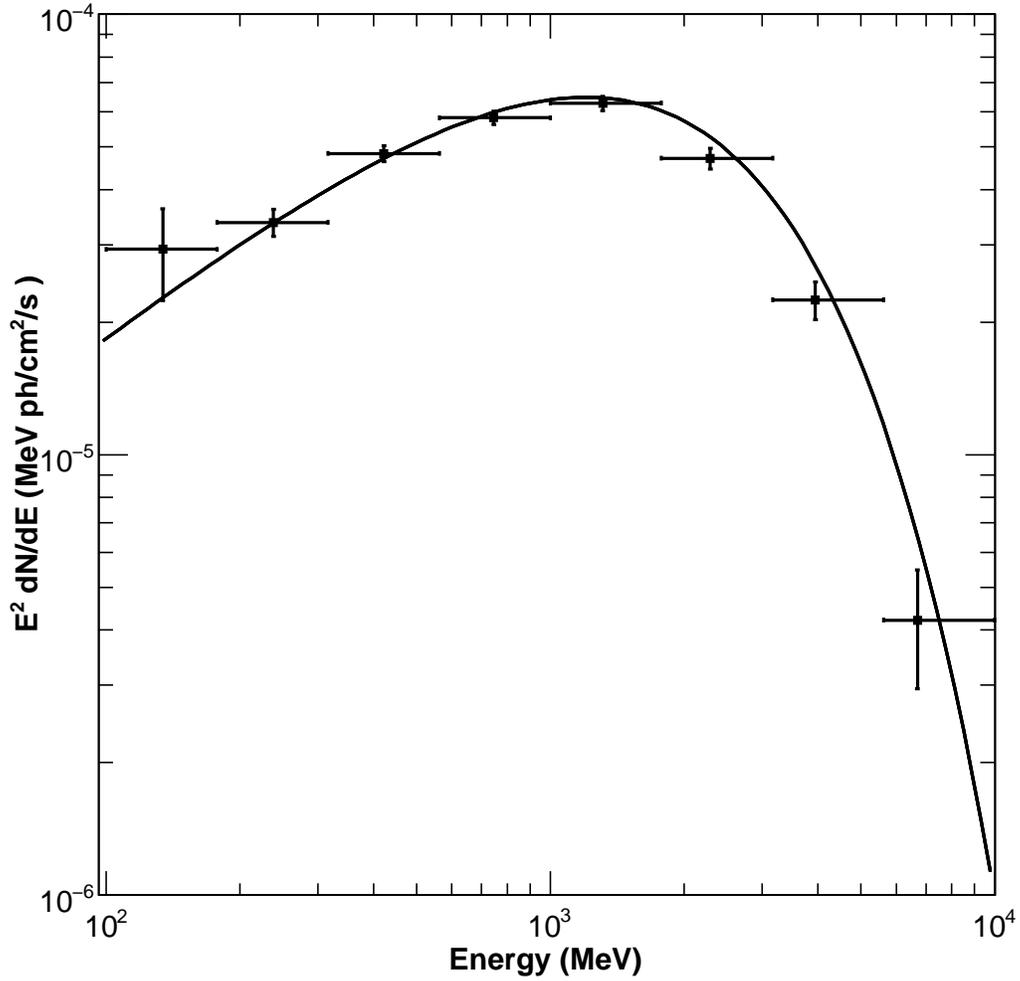}
 \caption{Gamma-ray spectrum for \PSROne. The curve represents the fit with simple-exponential cutoff($b=1$) shape in the full energy range of 100 MeV$-$100\,GeV. The spectrum points on the curve were obtained from an independent fit in each energy bin, with a model-independent method, as explained in the text. }
 \label{sed1055_acdc}
\end{figure}

\begin{figure}[ph]
 \centering
 \includegraphics[scale=0.8]{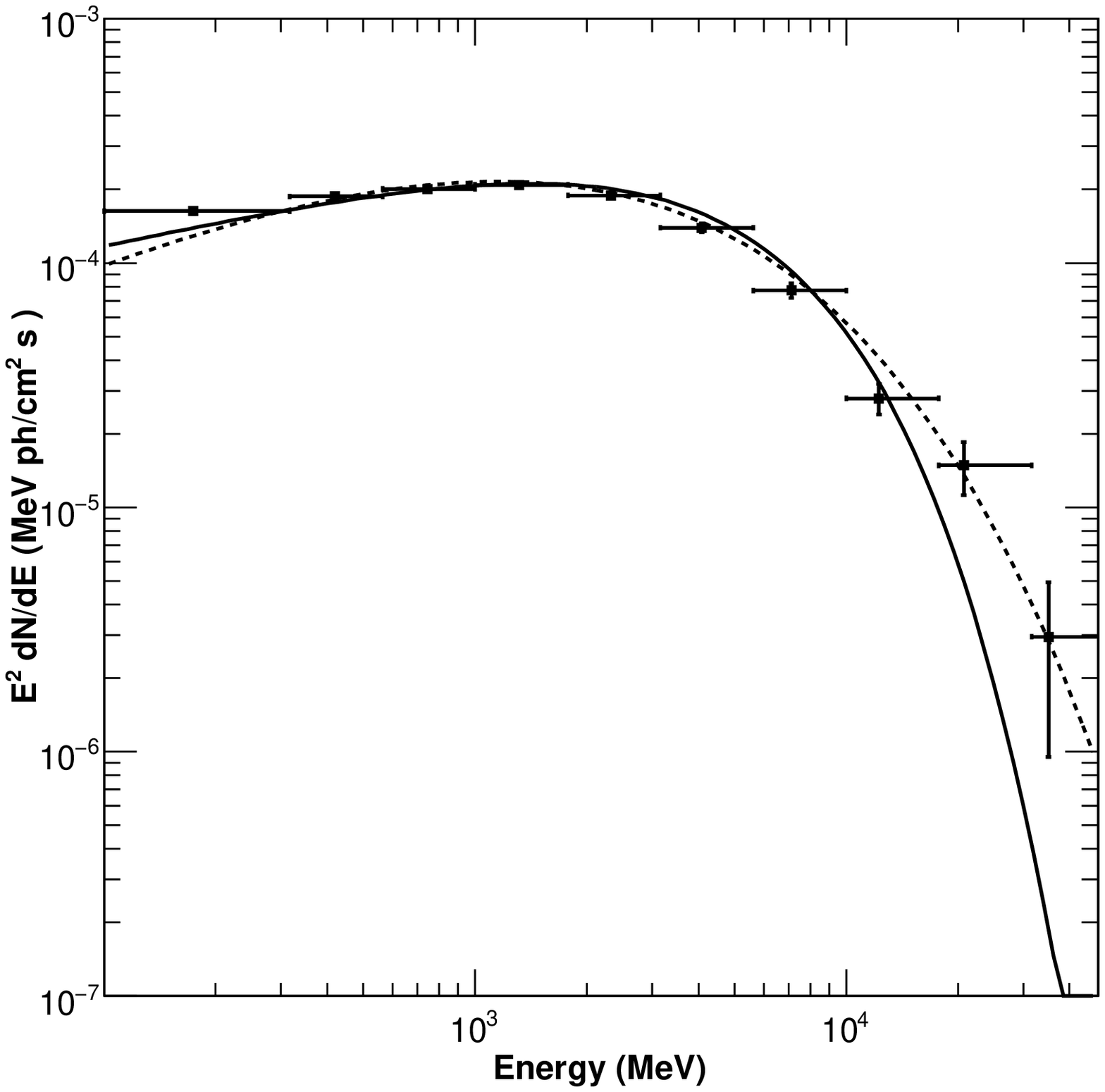}
 \caption{Gamma-ray spectrum  for \PSRTwo. The solid curve represents the fit with simple-exponential-cutoff ($b=1$) shape and the dashed curve represents the fit with gradual-exponential-cutoff ($b<1$) shape in the full energy range of 100 MeV- 100\,GeV. The spectrum points on the curve were obtained from an independent fit in each energy bin, with a model-independent method, as explained in the text.}
 \label{SedJ1709}
\end{figure}

\begin{figure}[ph]
 \centering
 \includegraphics[scale=0.8]{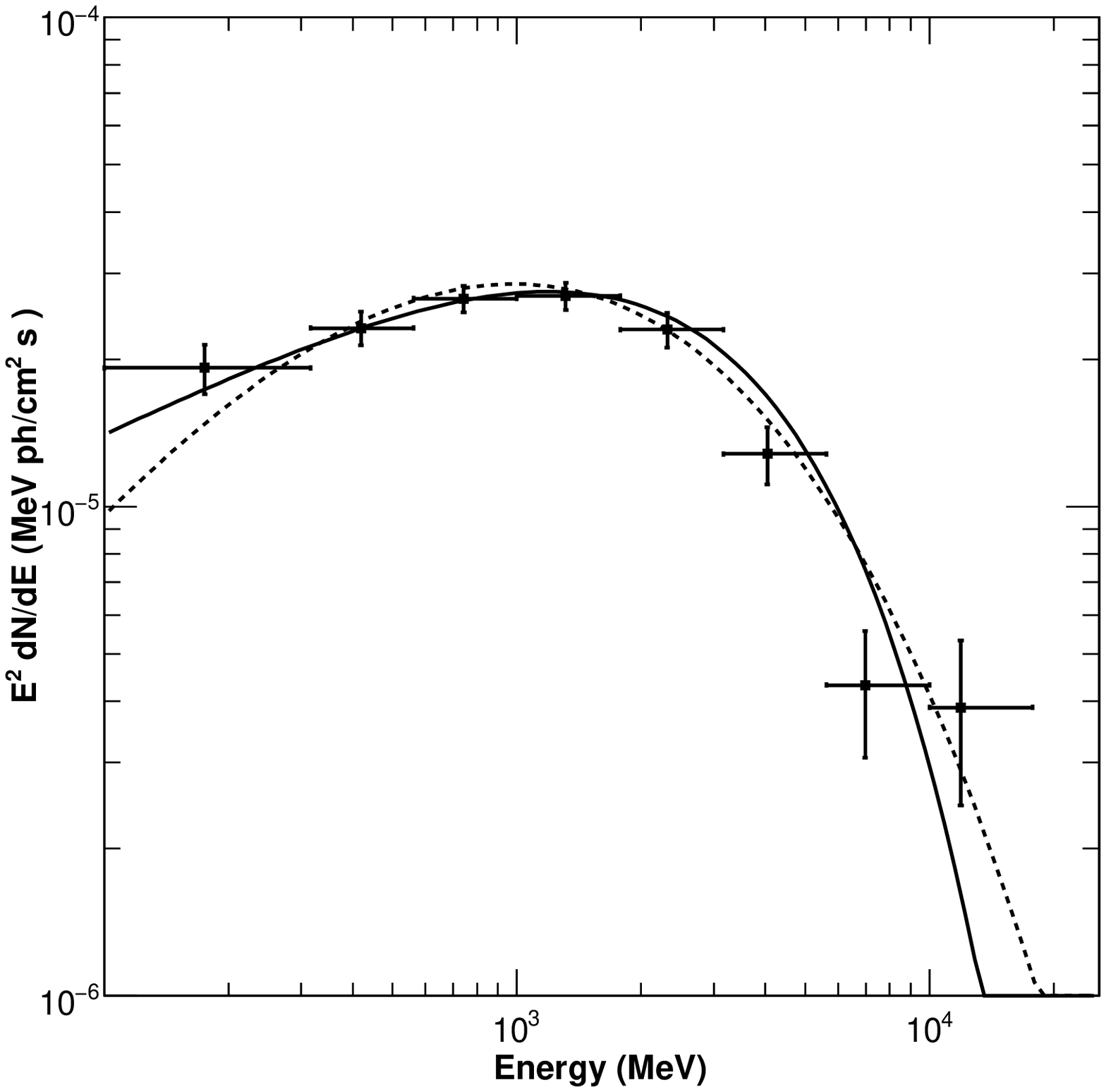}
 \caption{Gamma-ray spectrum  for \PSRThree. The solid curve represents the fit with simple-exponential-cutoff ($b=1$) shape and the dashed curve represents the fit with gradual-exponential-cutoff ($b<1$) shape in the full energy range of 100 MeV- 100\,GeV. The spectrum points on the curve were obtained from an independent fit in each energy bin, with a model-independent method, as explained in the text.
}
 \label{sed1952}
\end{figure}

In order to better evaluate the background in the region, we first fitted the diffuse background and nearby sources using the photons arriving in the off-pulse interval of the pulsars. The full-band phase-averaged spectrum for the pulsar of interest was then obtained using all the photons from the full phase range, and fixing the diffuse background and all nearby source spectral parameters to the best fit model obtained from the off-pulse analysis, properly rescaled to the full phase interval. At this step, the spectrum of the pulsar of interest was modeled with a power-law with exponential cutoff shape given by 
\begin{equation}
\frac{dN}{dE} = K E_{\rm GeV}^{-\Gamma}\exp \left[- \left(\frac{E}{E_{\rm cutoff}}\right)^{b} \right],
\label{expcutoff}
\end{equation}
where three parameters, the differential flux $K$ (ph cm$^{-2}$ s$^{-1}$ MeV$^{-1}$), the photon index $\Gamma$, and the cutoff energy, $E_{\rm cutoff}$, were allowed to be free in the fit and $b$ was fixed to be 1 (i.e., simple exponential). The fitted spectrum, with this assumed spectral model, for each pulsar is given in Table~\ref{Tbl:SpectResults} and plotted as the solid curve in Figures \ref{sed1055_acdc}, \ref{SedJ1709}, and \ref{sed1952} for these pulsars.

The presence of cutoffs in the spectra of the pulsars under study was checked by fitting each spectrum with a power-law shape, and this model was rejected significantly for all three pulsars in favor of the cutoff model by a likelihood ratio test. A broken-power-law spectrum was the best-fit model for the \PSRTwo\ with EGRET data. This spectrum model was also tested for \PSRTwo\ with the LAT data, and it was rejected by $43.4\sigma$ in favor of a simple exponential cutoff shape. 

Finally, the signal-to-noise ratios obtained from these pulsars were sufficiently high that we tested the validity of the simple-exponential-cutoff shape assumption by letting the $b$ parameter free in the spectral fit model. We found that for \PSRTwo\ and \PSRThree\ the best-fit spectrum has $b<1$, rejecting the $b=1$ shape with $7.2\sigma$ for \PSRTwo\ and with $2.9\sigma$ for \PSRThree. The fit for \PSROne\ did not improve with a free $b$ parameter and resulted in a value for $b$ consistent with 1. The spectral parameters obtained from the fits with a free $b$ parameter are also given in Table~\ref{Tbl:SpectResults} and the $b<1$ fits are shown with the dashed curves in Figures \ref{SedJ1709} and \ref{sed1952}.

The flux points in Figures~\ref{sed1055_acdc}, \ref{SedJ1709}, and \ref{sed1952} were obtained with an analysis method that is independent of the model used to fit the sources in the region. We divided the data set into energy bins logarithmically spaced and in each of them we applied the \textit{gtlike} tool, but this time, for all the point-like sources, we assumed a power law spectrum with a fixed photon index of 2, and a free flux parameter. We also modeled the Galactic diffuse emission and the isotropic component as before. In this way, we obtained the flux values for all the sources in each energy bin assuming that in the energy bin the spectral shape can be safely approximated with a power law. This method is more accurate as the energy bin size is reduced, but statistical considerations limit the minimum bin width.

From the comparison of the spectral flux points and the full-band spectrum curve, it is clearly seen that while a simple-exponential-cutoff ($b=1$, solid curve) shape is a good fit for \PSROne, it does not fit the high energy points properly for \PSRTwo\ or \PSRThree. A gradual-exponential-cutoff ($b<1$, dashed curve) model for these two pulsars agrees very well with all the points up to highest energies, consistent with the preference for this model over a $b=1$ model in the full-band fit. A best-fit phase averaged spectral shape with $b<1$ is similar to what has been found for the Vela pulsar spectrum~\citep{LATVela2}. Although a power law with a gradual cutoff, $b<1$, does not have any physical meaning, it is a possible indication for large variations in $E_{\rm cutoff}$ with phase. The phase-averaged spectrum can be a combination of several spectra exhibiting simple exponential cutoff ($b=1$) with different cutoff energies. The same behavior was demonstrated in Vela using simulations~\citep{LATVela2}.

\subsubsection{Phase-resolved Spectrum}
\label{Sec:PRSpect}
The large number of photons collected by the {\it Fermi} LAT from these pulsars allows us to obtain phase-resolved spectra in order to study the variation of the spectral parameters with phase. The phase bins for the phase-resolved spectra were defined such that in each bin a fixed-count of 500 photons for \PSROne, 1500 photons for \PSRTwo, and 1000 photons for \PSRThree\ were collected in the energy-dependent radius defined in Section~\ref{Sec:PlsProfile}. The spectral analysis data set covering the 10\degree\ ROI around each pulsar was then divided into these phase bins, and an independent spectral fit was performed for each phase bin. For each spectral fit, the diffuse background and nearby sources were modeled with the same best-fit model obtained from the off-pulse analysis, thus only variations from bin to bin were due to the pulsar. The spectral fit parameters were obtained in each phase bin where the significance of the pulsar emission was above the $5\sigma$ level. 

\begin{figure}[ph]
 \centering
 \includegraphics[width=0.85\textwidth]{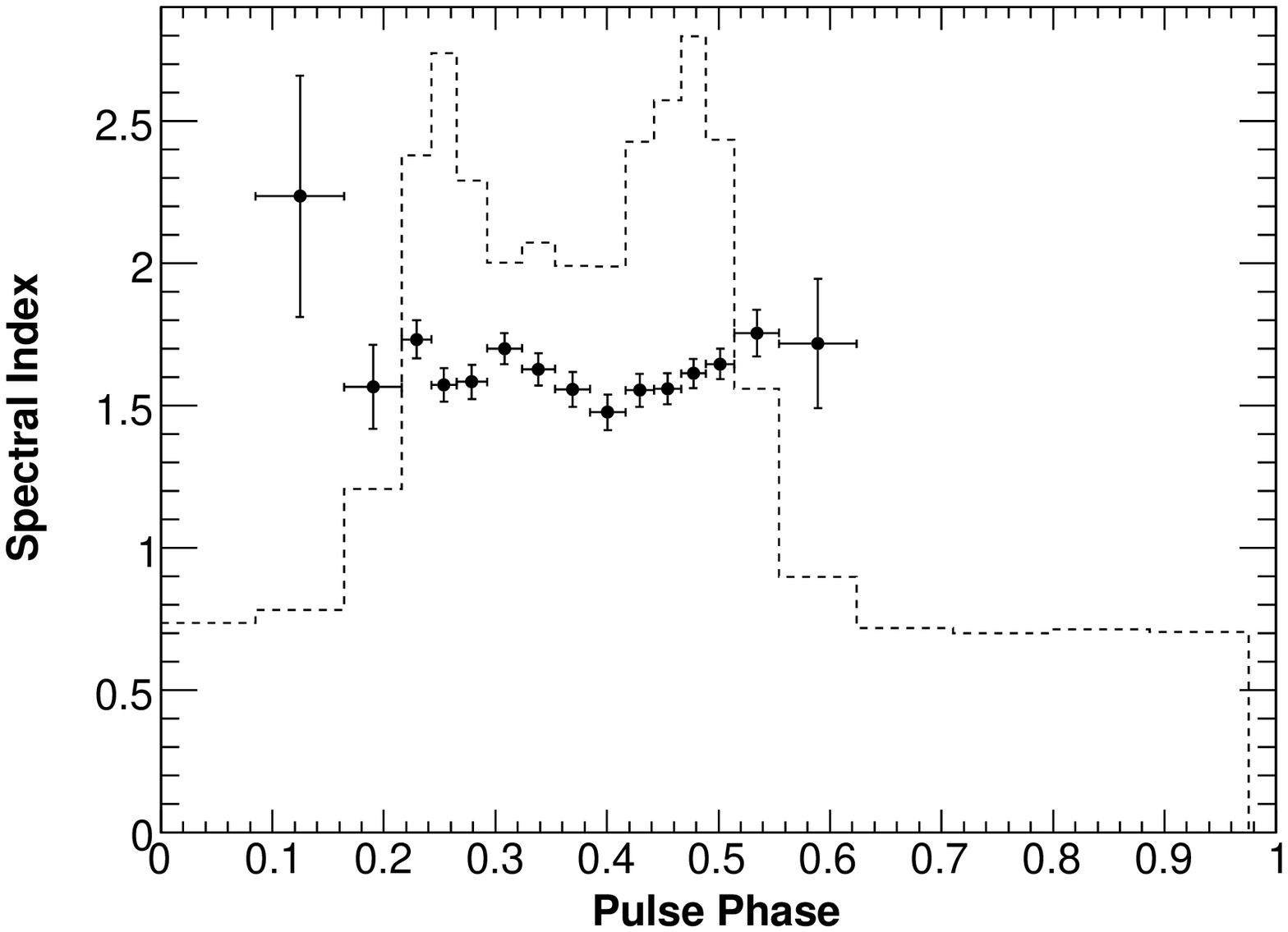}
 \includegraphics[width=0.85\textwidth]{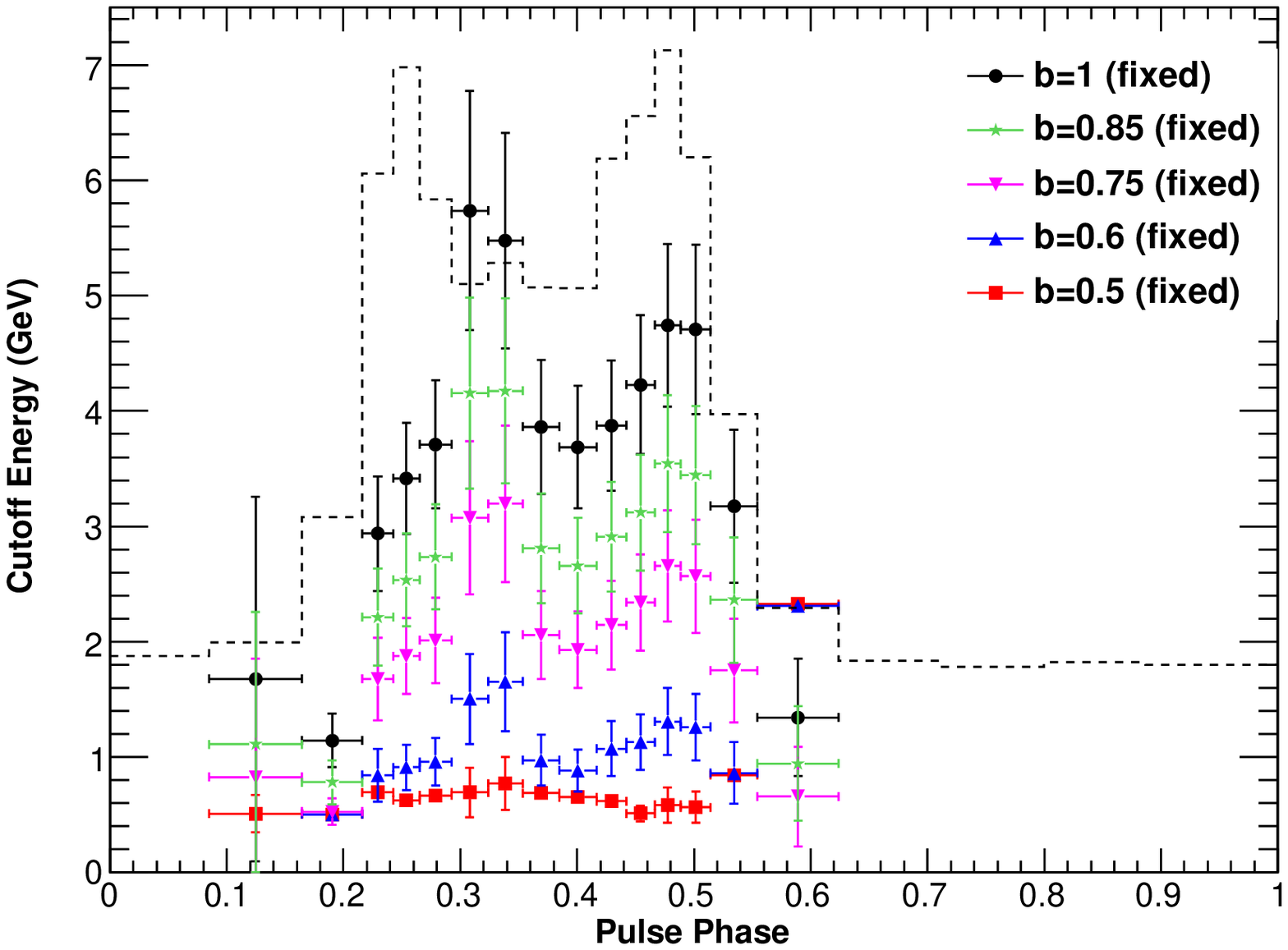}
 \caption{Evolution of the photon index (top panel) and the cutoff energy (bottom panel) of the pulsed emission spectrum through the pulse profile of the \PSRTwo.}
 \label{Fig:J1709PhRslvd}
\end{figure}

Although the data in each phase bin were not adequate for any of the pulsars to make a fit to test the shape of the cutoff while allowing $b$ to vary, additional fits for \PSRTwo, the brightest of these three pulsars, were performed in each phase bin fixing $b$ to four different values between 0.5 and 1.0, and it was found that none of those gradual-cutoff-shape models gave a significantly better fit in any phase bin than the model with simple-exponential-cutoff shape ($b=1$), with the current statistics, as opposed to the case of the phase-averaged spectra. The comparison of the cutoff energies obtained for each fixed $b$ value also revealed that the $E_{\rm cutoff}$ and $b$ parameters in Equation~(\ref{expcutoff}) are strongly correlated (see Figure\ref{Fig:J1709PhRslvd}), as also evident in the phase-averaged spectrum results with $b$ fixed at 1 or left free (Table~\ref{Tbl:SpectResults}). Thus, for these limited-counting-statistics spectra where we can not obtain a significantly better fit in any phase bin among different values of $b$,  we assumed a shape of a power law with simple-exponential cutoff ($b=1$) for each pulsar, allowing only the K, $\Gamma$, and $E_{\rm cutoff}$ parameters to vary in the fit.

Figures~\ref{B1055_PhaseScan_Plots}, \ref{Fig:J1709PhRslvd} and \ref{phaseresolved_j1952}  show the variation of the spectral parameters with phase for each pulsar, while Tables~\ref{B1055_PhaseScan_Table}$-$\ref{tab:j1952interval} report the spectral parameters obtained from the fit in each phase bin for each pulsar. The quoted fluxes are normalized by the width of the phase bin.

The spectrum of \PSROne\ was measured in eight bins in the phase range $\varphi= 0.28 - 0.64$. No strong correlations between spectral parameters and phase are observed in this region. More data will be needed to fully characterize any dependence of the spectral parameters with phase.

The spectrum of \PSRTwo\ was measured in 15 bins in the phase range $\varphi= 0.085-0.624$. The photon index varies slowly throughout the pulse profile, with a minimum around the leading edge of P2 and increasing toward the tails of the pulsed emission region. The cutoff energy shows a more drastic evolution, changing from a minimum of $1.14 \pm 0.23$\,GeV at the leading edge of P1 to a maximum of $5.74 \pm 1.04$\,GeV around the location of P3 seen in the pulse profile above 3\,GeV. The cutoff energy has an increasing trend from P1 to P2, but has a sharp maximum near P3, and falls off at the trailing edge of P2. 

Lastly, the spectrum of \PSRThree\  was measured in five bins at the locations of the peaks. The photon index is quite constant over the two peaks, while the cutoff energy has an increasing trend from P1 to P2.

Although more data are necessary to clearly assess any trends, the evolution of cutoff energy with phase for these pulsars seems to be compatible with the prediction that the $b<1$ preference in the phase-averaged spectrum results from large variations of cutoff energy with phase, while $b$ consistent with 1 implies that the cutoff energy should be constant with phase. The cutoff energy for \PSROne\ is quite constant over the phase, and the $b=1$ cutoff shape is a sufficiently good fit for the phase-averaged spectrum. On the other hand, \PSRTwo\ shows the most drastic evolution of the cutoff energy over its phase and the best-fit phase-averaged spectrum of this pulsar requires the lowest value of $b=0.506 \pm 0.021_{\rm stat} \pm 0.035_{\rm sys}$.

\subsection{Search for Off-pulse Emission}
\label{Sec:OffEm}

The off-pulse phases of emission from these pulsars were also analyzed to search for potential unresolved nebular sources around the pulsars. For these searches, a minimum emission phase range in the off-pulse regions was selected to avoid contamination from the pulsed emission peaks.  Spectral analyses, as described in Section~\ref{Sec:SpectAn}, were performed in these off-pulse phase ranges by first assuming no steady emission exists at the location of the pulsar (null hypothesis) and then assuming a point source with a power-law emission spectrum at the pulsar locations (test hypothesis). The log-likelihood values of both fits for each pulsar are compared to calculate the test statistic (TS) of a possible off-pulse source at the location of the pulsar from the difference between the log likelihood with ($L_1$) and without ($L_0$) the source, i.e., TS = 2($L_1 - L_0$).

We selected photons from the off-pulse region $(\varphi= 0.0-0.25)$ and $(\varphi= 0.65-1.0)$ to search for nebular emission from \PSROne. From the spectral fit performed with \textit{gtlike}, we cannot claim any clear detection of a nebula with a TS value of 15. We set an upper limit at 95\% confidence level for the flux from the nebula to $F(>100 {\rm MeV})= 6.67 \times 10^{-8}$ ph cm$^{-2}$s$^{-1}$.

For \PSRTwo, we selected an off-pulse phase region of $(\varphi=0.0-0.1)$ and $(\varphi= 0.7-1.0)$ to search for a steady emission. We did not detect any significant point-like emission in that off-pulse phase interval with a TS of 23.4 for a test source at the location of this pulsar. We set an upper limit at 95\% confidence level for the flux from the nebula of $F(>100 {\rm MeV})= 8.51 \times 10^{-8}$ ph cm$^{-2}$s$^{-1}$. This result is consistent with the non-detection of a point-like off-pulse source by EGRET~\citep{Thompson1996} and HESS~\citep{Aharonian2005-1} around this pulsar. The analysis to search for extended emission around this pulsar is ongoing using the LAT data in the light of detection of a significant extended emission by HESS at an offset of 0.5\degree\ from the pulsar.

The off-pulse emission from \PSRThree\ has been investigated in the phase interval of $\varphi= 0.75-1.00$.  We did not detect any significant point-like emission with a TS of 2.8 for a test source at the location of this pulsar. Thus, no emission is detected from CTB80, and we calculated an upper limit of $F(>100 {\rm MeV})= 6.5 \times 10^{-8}$ ph cm$^{-2}$s$^{-1}$ at 95\% confidence level.

\section{Discussion}
\label{Discussion}

The superior sensitivity and high statistics obtained with the LAT data allow construction of light curves and spectra with better precision,  providing tighter observational constraints on the pulsed emission models than previously possible. Currently, there are two classes of high-energy pulsed emission models: the PC~\citep{Daugherty1996} models predicting a lower altitude emission from near the magnetic poles of the neutron star and high-altitude emission models like OG~\citep{Romani1996} or two-pole caustic (TPC)~\citep{Dyks2003} models, like slot gap (SG)~\citep{Muslimov2004}, which predict emission higher in the magnetosphere extending up to the light cylinder.

The on-pulse regions of the light curves cover a wide range in phase, from 30\% in the case of \PSRThree\ to 46\% for \PSRTwo, while the radio emission peaks in a very narrow angle for all of them. This feature suggests that the gamma-ray beams cover a large solid angle compared to the radio emission and favors the prediction of high-altitude emission models, such as TPC and OG.

The maximum energy $\epsilon_{\rm max}$ of the pulsations provides a lower limit to the altitude of gamma-ray emission, since it must lie below the $\gamma-B$ magnetic pair creation absorption threshold. This bound can be simply expressed in terms of the surface polar magnetic field strength $10^{12}B_{12}$G, the pulsar period $P$ (in seconds), and $\epsilon_{\rm max}$, inverting Equation~(1) of \citet{baring04}: 
\begin{equation}
r\gtrsim(\epsilon_{\rm max} B_{12}/1.76\hbox{GeV})^{2/7}\, P^{-1/7}\, R_{\ast}
\end{equation}
where $R_{\ast}$ is the radius of the neutron star. The good photon statistics permit tracing the spectrum deep into the turnover for each of these pulsars: a super-exponential $\gamma-B$ pair attenuation feature is excluded by the spectra at energies above the cutoff energy $E_{\rm cutoff}$ specified in Equation~(\ref{expcutoff}). Accordingly, it is safe to adopt a value of around $2E_{\rm cutoff}$ for the maximum energy $\epsilon_{\rm max}$ in bounding the emission altitude.  If one chooses $\epsilon_{\rm max}=3.0$ GeV in \PSROne\ ($E_{\rm cutoff}\approx 1.50$ GeV), this yields $r\gtrsim 1.5 R_{\ast}$. For \PSRTwo, the higher energy choice of $\epsilon_{\rm max}=8.9$ GeV yields the bound $r\gtrsim 3.0 R_{\ast}$.  For \PSRThree, the limit is $r\gtrsim 1.8 R_{\ast}$ when adopting $\epsilon_{\rm max}=5.6$ GeV for $E_{\rm cutoff}\approx 2.8$ GeV.  Clearly these bounds exclude emission very near the stellar surface, adding support for an SG or OG acceleration locale in these pulsars.

The gamma-ray efficiency, $\eta_{\gamma}$, of the pulsar can be obtained from the relation, $\eta_{\gamma}=L_{\gamma}/\dot{E}$, where $L_{\gamma}$ is the total gamma-ray luminosity given as
\begin{equation}
L_{\gamma} = 4\pi f_{\Omega}(\alpha,\zeta_{E})\,d^{2}\,G_{100}.
\label{LGamma}
\end{equation}
In this equation, $G_{100}$ is the observed energy flux above 100 MeV at the Earth line of sight (at angle $\zeta_E$ to the rotation axis), $d$ is the pulsar distance, and $f_{\Omega}$ is the beaming correction factor that depends on the geometry of the emission. For the PC model~\citep{Daugherty1996}, the gamma-ray emission originates at a few stellar radii from the surface, implying an emission with a small solid angle, that is, $f_{\Omega} \ll$ 1. For both OG and SG models, where the emission is far away from the neutron star, the resulting $f_{\Omega}$ values can be near or even greater than 1. Aside from the unknown factor $f_{\Omega}$, a large uncertainty on the gamma-ray efficiency arises from the determination of the distance. The distances obtained from the literature for the three pulsars under study are listed in Table~\ref{Tbl:PSRpars} and the observed energy flux values, $G_{100}$, from these pulsars are listed in Table~\ref{Tbl:SpectResults}.

Using these parameters in Equation~(\ref{LGamma}), we found $L_{\gamma}=(1.7 \pm 0.7) \times 10^{34}\,f_{\Omega}$\,erg\,s$^{-1}$ and $\eta_{\gamma}=(0.58 \pm 0.23)\,f_{\Omega}$ for \PSROne. A similar calculation for \PSRThree\ leads to $L_{\gamma}=(6.9 \pm 2.9) \times 10^{34}\,f_{\Omega}$\,erg\,s$^{-1}$ and an efficiency of $\eta_{\gamma}=(0.019 \pm 0.008)\,f_{\Omega}$. Finally, for \PSRTwo\ the $L_{\gamma}$ is found in the range $(2.58 \times 10^{35}\,f_{\Omega}$ $ - 1.70 \times 10^{36}\,f_{\Omega}$) given the range of its estimated distances and its gamma-ray efficiency is in the range $(0.076\,f_{\Omega} - 0.50\,f_{\Omega})$. From the EGRET observations, there was a noted trend that gamma-ray pulsar efficiency scales as $\dot E^{1/2}$~\citep{Thompson1999}, with a physical origin earlier predicted by \citet{Harding1981}.  {\it Fermi} pulsars have confirmed this trend, at least for $\dot E > 10^{34}$, that $\eta_{\gamma} \simeq 0.034 (\dot E / 10^{36}\,\rm erg\,s^{-1})^{-1/2}$~\citep{LATPULSARS}.  \PSRThree\ follows this relation almost exactly, while the efficiency of \PSROne\ lies above by a factor of 2, and the efficiency range of \PSRTwo\ lies substantially above the efficiency predicted with this relation.

The number of main peaks and the peak separation may be used to constrain the geometry of the emission for high-altitude models. The measurements of the viewing angle, $\zeta$, and the magnetic inclination angle, $\alpha$, obtained from the torus fits to the X-ray images \citep{Ng2008} or from polarization measurements~\citep{Weltevrede2009}, listed in Table~\ref{Tbl:MeasPars}, when available, provide additional constraints on the model, as described in \citet{Romani2009}. The possible ranges of geometrical parameters, $\alpha$ and $\zeta$, can be obtained from the allowed phase space of the predicted geometry of the two geometrical models constrained with the measurements. The light curves modeled for this geometry are then compared with the observed light curves to determine a goodness of fit, $\chi$. The inferred angles, $\alpha$ and $\zeta$, and the goodness of the fits, $\chi$, for the TPC and OG models, obtained from this procedure \citep{Romani2009}, are listed in Table~\ref{Tbl:AtlasPars}.  These geometrical considerations might give hints about the underlying emission model.

\citet{Romani2009} find that, in some cases, several models can produce acceptable fits. PSR J1952+3252 has only limited X-ray/radio constraints and both TPC models and OG models can produce light curves with comparable fit statistics. In the case of PSR J1709-4429, fits consistent with the externally measured $\alpha$ and $\zeta$ can be found for both TPC and OG, although the fit statistic for the latter is substantially better. For PSR J1057-5226 no viable TPC models are present anywhere near the radio-determined $\alpha$ and $\zeta$, while the OG produces reasonable light curves for these values. 

The predictions for the pulsed emission geometry in these models assume the efficiency relation $(w =) \eta \simeq (10^{33}/\dot{E})^{0.5}$ of \citet{watters09}, and the gamma-ray efficiencies (or gap widths) derived from this assumption are also listed in Table~\ref{Tbl:AtlasPars}. Combining these assumed gamma-ray efficiency values with the gamma-ray efficiencies derived from the observed gamma-ray luminosities, one can derive a flux correction factor $f_{\Omega}$ for each pulsar and compare this value against the predicted flux correction factor for each pulsar from these models. For \PSRThree\ the flux correction factor $f_{\Omega}$, derived from this relation, is compatible with the predictions of both models. However, the derived flux correction factor $f_{\Omega}$ for the other two pulsars, \PSROne\ and \PSRTwo, are significantly lower than the predictions of the geometrical models. Thus, those predictions of geometrical models for flux correction factors must be revised. In particular, \PSRTwo\ does appear to have a substantially higher efficiency than expected for its estimated distance and predicted beaming.

The phase lag between the radio pulse and the first gamma-ray pulse is listed in the third column of Table~\ref{Tbl:MeasPars}. These are larger than expected from a basic geometric model of both OG and two-pole models like the ones discussed in \citet{watters09}. The main reason for the difference is their underlying assumption that the radio emission comes from the PC region near the surface. However, if an additional assumption of high-altitude radio emission is invoked as discussed in \citet{Karastergiou2007}, \citet{watters09}, and \citet{Venter2009}, the radio pulse will arrive at earlier phase because of aberration, and retardation due to the finite-time-of-flight effect, and the larger resulting phase lag with the gamma-ray pulse better matches the observed ones.

The phase-resolved spectral measurements that are now achievable with the LAT will provide another important diagnostic for model comparisons.  Among the three pulsars analyzed in this paper, only \PSRTwo\ so far has measurements with small enough errors and in fine enough phase bins to show significant variations.  The variations seen for \PSRTwo\ in photon index and cutoff energy show patterns that are similar to those measured for Vela~\citep{LATVela2} and Geminga~\citep{LATGeminga}, notably a slight hardening in index in the bridge region and a rise to maximum values of the cutoff energy in the peaks.  In general, the stronger variation of cutoff energy with phase may well be tied to variations in emission radius and/or magnetic-field-line curvature since most models predict that the emission at GeV energies is due to radiation-reaction-limited curvature radiation with exponential cutoffs in the range of the measured cutoff energies.

\section{Conclusion}

A common thread that links \PSROne, \PSRTwo, and  \PSRThree\ is that they were all detected by EGRET, although not nearly as strongly as Vela, Crab, and Geminga.  One question is why these three were visible to EGRET when we now know there are at least 46 gamma-ray pulsars~\citep{LATPULSARS}, a number of which are in EGRET unidentified source error boxes. These three were not among the brightest Galactic sources seen by EGRET; in fact, PSR~J1952+3252 did not appear in the third EGRET catalog, being detected only by its pulsations. The basic answer appears to be that these were among the brightest of the gamma-ray pulsars that had good radio timing in the EGRET era. Three other pulsars that had good radio timing were seen by EGRET as marginal detections (now all confirmed by LAT observations): PSR~J0218+4232~\citep{Kuiper2000}, PSR~B0656+14~\citep{Ramanamurthy1996} and PSR~B1046$-$58~\citep{Kaspi2000}. PSR~B0740$-$28 had good timing in the EGRET era, but the LAT detection~\citep{LATPULSARS} shows that the gamma-ray flux is well below the EGRET sensitivity.  Many of the radio pulsars that have been detected by {\it Fermi} had timing programs started only after the primary EGRET mission was over. Therefore, it seems clear that \PSROne, \PSRTwo, and \PSRThree\ are not fundamentally different from the other LAT-detected pulsars.

The other key question addressed by this paper is whether the unique features suggested by the EGRET observations are supported by the LAT observations.  The answer is mixed.

\PSROne\  in the EGRET era appeared to have a very high efficiency, high enough to argue against models with large beams as being physically unrealistic.  Improved distance estimates now put this pulsar closer to Earth, but its efficiency still appears to be higher than that of most gamma-ray pulsars.

\PSRTwo\  was unique among the EGRET pulsars in having an energy spectrum well fit by a broken power law, rather than a single power law or a cutoff at high energy. This fit was driven by the indications of significant pulsed emission above 10 GeV, a feature that is confirmed by the LAT data.  Although the exponential cutoff spectral form fits the LAT data, it requires a high-energy cutoff and it was not dramatically better than a broken power-law fit until more than 9 months of data were included in the analysis.  However, the LAT best fit to the phase-averaged spectrum yields $b < 1$ so that the spectrum falls off more slowly than a pure exponential, as may be expected from a cutoff whose energy varies with pulsar phase.  This shape may have allowed an EGRET fit with a broken power law.

\PSRThree\  showed no evidence of a spectral cutoff in the EGRET data, extending out to the 10$-$30 GeV energy band.  The LAT results show a clear exponential cutoff at a much lower energy, so this feature from the EGRET era is not confirmed.  As noted, however, this pulsar had very limited statistics with EGRET (only two photons above 10 GeV).  \PSRThree\  does not stand out among the LAT pulsars in any respect.

\PSROne\ and \PSRTwo\ are among the minority of LAT pulsars that do not show widely spaced double pulses.  Several pulsars in the first LAT catalog have closely spaced double pulses similar to these, notably PSR~J0007+7303 (the CTA1 pulsar)~\citep{CTA1} and PSR~J1509$-$5850~\citep{PSR1509m5850}.  So although these shapes are not the most common, they should be predicted by successful outer magnetosphere models.  Continued study will be needed in order to derive detailed information about the pulsar magnetospheres for these two.


\acknowledgments

The \textit{Fermi} LAT Collaboration acknowledges generous ongoing support from a number of agencies and institutes that have supported both the development and the operation of the LAT as well as scientific data analysis. These include the National Aeronautics and Space Administration and the Department of Energy in the United States, the Commissariat \`a l'Energie Atomique and the Centre National de la Recherche Scientifique / Institut National de Physique Nucl\'eaire et de Physique des Particules in France, the Agenzia Spaziale Italiana and the Istituto Nazionale di Fisica Nucleare in Italy, the Ministry of Education, Culture, Sports, Science and Technology (MEXT), High Energy Accelerator Research Organization (KEK) and Japan Aerospace Exploration Agency (JAXA) in Japan, and the K.~A.~Wallenberg Foundation, the Swedish Research Council and the Swedish National Space Board in Sweden.

Additional support for science analysis during the operations phase is gratefully acknowledged from the Istituto Nazionale di Astrofisica in Italy and the Centre National d'\'Etudes Spatiales in France.

This work made extensive use of the ATNF pulsar  catalog \citep{ATNFcatalog}\footnote{http://www.atnf.csiro.au/research/pulsar/psrcat}.

The Parkes radio telescope is part of the Australia Telescope which is funded by the Commonwealth of Australia for operation as a National Facility managed by the CSIRO.

The Green Bank Telescope is operated by the National Radio Astronomy Observatory, a facility of the National Science Foundation operated under cooperative agreement by Associated Universities, Inc.

The Jodrell Bank observatory is operated by the Science and Technology Facilities Council of the United Kingdom.

The Nan\c cay Radio Observatory is operated by the Paris Observatory, associated with the French Centre National de la Recherche Scientifique(CNRS).

\bibliographystyle{apj}
\bibliography {apj-jour,3EGRETPulsars_v1.7}


\newpage

\begin{table*}[tb]
\begin{center}
\begin{tabular}{llcccccc}
\hline \noalign{\smallskip}
Pulsar & BName & $P$    & $\dot{P}$      & $\dot{E}$ & Age $^{(a)}$  & $B_{\rm{surface}}$ & Distance \\
       &       & (ms) & ($10^{-15}$) & $10^{34}$ ergs\,s$^{-1}$ & ($10^3$ yr) & ($10^{12}$ G) & (kpc) \\[3pt]
\hline \noalign{\smallskip}
\PSROne\ & B1055$-$52 & 197  & 5.83 & 3.01 & 535  & 1.09  & $0.72 \pm 0.2^{(b)}$ \\
\PSRTwo\ & B1706$-$44 & 102  & 93.0 & 341  & 17.5 & 3.12  & $1.4 - 3.6^{(c)}$ \\
\PSRThree\ & B1951$+$32 & 39.5 & 5.84 & 374  & 107  & 0.486 & $2.0 \pm 0.6 ^{(d)}$ \\
\hline
\end{tabular}
\end{center}
\tablenotetext{a}{Characteristic age: $P/2\dot{P}$}
\tablenotetext{b}{Estimated using dispersion measure~\citep{Cordes2002, Weltevrede2009}.}
\tablenotetext{c}{\citet{Koribalski1995, McGowan2004}}
\tablenotetext{d}{Evaluated from the kinematic model~\citep{Greidanus1990}}
\caption{Various Characteristic Parameters for the Three Pulsars. Values are taken from the ATNF database~\citep{ATNFcatalog} except the distance estimates, for which the references are listed separately. We adopt the same distance estimates used for the first {\it Fermi} pulsar catalog~\citep{LATPULSARS}.}
\label{Tbl:PSRpars}
\end{table*}

\begin{table*}[tp]
\vspace{0.2cm}
\begin{center}
\begin{tabular}{lcccr}
  \hline
  Pulsar       & Observer        & Validity Period (MJD) & TOA & rms \\
   \hline \hline\noalign{\smallskip}
  \PSROne\ *   & Parkes          & 54220.2$-$55071.9 & 47 & 62\,$\mu$s   \\[1pt]
  \hline\noalign{\smallskip}
  \PSRTwo\     & Parkes          & 54220.6$-$55072.3 & 49 & 1255\,$\mu$s \\
  \PSRTwo\ *   & {\it Fermi} LAT & 54647.4$-$55074.6 & 23 & 378\,$\mu$s  \\[1pt]
  \hline\noalign{\smallskip}
  \PSRThree\ * & Nan\c cay       & 54607.1$-$55075.9 & 17 & 228\,$\mu$s   \\[1pt]
  \hline 
\end{tabular}
\end{center}
\caption{Timing Observations. The number of TOAs used to derive the timing solutions and the post-fit root-mean-square (rms) of the timing solution are also given. The (*) denotes the observations that give the timing solution used in this analysis.}
\label{Tbl:PSRTimeSol}
\end{table*}

\begin{table*}[tp]
\vspace{0.2cm}
\begin{center}
\begin{tabular}{l|cccc}
  \hline\noalign{\smallskip}
   Energy Band     & P1 Location     & P1 HWHM         & P2 Location     & P2 HWHM \\
   \hline\noalign{\smallskip}
   0.1 $-$ 100 GeV &  0.31$\pm$0.01  &  0.04$\pm$0.01  &  0.59$\pm$0.01  &  0.03$\pm$0.01 \\
   $>$3.0 GeV      &  0.32$\pm$0.02  &  0.04$\pm$0.03  &  0.59$\pm$0.02  &  0.04$\pm$0.02 \\
   1.0 $-$ 3.0 GeV &  0.32$\pm$0.01  &  0.03$\pm$0.01  &  0.59$\pm$0.01  &  0.02$\pm$0.01 \\
   0.3 $-$ 1.0 GeV &  0.31$\pm$0.01  &  0.03$\pm$0.01  &  0.59$\pm$0.01  &  0.03$\pm$0.01 \\
   0.1 $-$ 0.3 GeV &  0.31$\pm$0.01  &  0.05$\pm$0.01  &  0.58$\pm$0.02  &  0.04$\pm$0.02 \\
\hline
\end{tabular}
\end{center}
\caption{The location and Gaussian half-width-at-half-maximum (HWHM) values for the leading and trailing peaks (P1 and P2) of the \PSROne\ pulse profiles in every energy band. These parameter values were obtained by fitting the peaks with Gaussian functions over a constant background.}
\label{Tbl:J1057PlsFit}
\end{table*}

\begin{table*}[tp]
\vspace{0.2cm}
\begin{center}
\begin{tabular}{l|cccccc}
  \hline\noalign{\smallskip}
   Energy Band   & P1 Location     & P1 HWHM (Outer)    & P1 HWHM (Inner)     \\
   \hline\noalign{\smallskip}
0.1 $-$ 100 GeV  & 0.242$\pm$0.002 & 0.027$\pm$0.002  &  0.072$\pm$0.008  \\
$>$10.0 GeV      &        $-$      &       $-$        &       $-$         \\
3.0 $-$ 10.0 GeV & 0.240$\pm$0.004 & 0.008$\pm$0.003  &  0.101$\pm$0.044  \\
1.0 $-$ 3.0 GeV  & 0.243$\pm$0.003 & 0.020$\pm$0.002  &  0.103$\pm$0.016  \\
0.3 $-$ 1.0 GeV  & 0.243$\pm$0.003 & 0.028$\pm$0.003  &  0.072$\pm$0.016  \\
0.1 $-$ 0.3 GeV  & 0.238$\pm$0.004 & 0.029$\pm$0.005  &  0.061$\pm$0.012  \\
\noalign{\smallskip}\hline\hline\noalign{\smallskip}
   Energy Band   & P2 Location      & P2 HWHM (Inner)    & P2 HWHM (Outer) \\
   \hline\noalign{\smallskip}
0.1 $-$ 100 GeV  &  0.492$\pm$0.003 &  0.118$\pm$0.013 & 0.032$\pm$0.003 \\
$>$10.0 GeV      &  0.501$\pm$0.009 &  0.037$\pm$0.013 & 0.009$\pm$0.010 \\
3.0 $-$ 10.0 GeV &  0.489$\pm$0.005 &  0.126$\pm$0.032 & 0.017$\pm$0.003 \\
1.0 $-$ 3.0 GeV  &  0.496$\pm$0.003 &  0.140$\pm$0.066 & 0.023$\pm$0.002 \\
0.3 $-$ 1.0 GeV  &  0.488$\pm$0.005 &  0.115$\pm$0.018 & 0.037$\pm$0.004 \\
0.1 $-$ 0.3 GeV  &  0.496$\pm$0.007 &  0.098$\pm$0.030 & 0.032$\pm$0.006 \\
\hline
\end{tabular}
\end{center}
\caption{The location and Lorentzian half-width-at-half-maximum (HWHM) values for the first and second peaks (P1 and P2) of the \PSRTwo\ pulse profiles in every energy band. These parameter values were obtained fitting the peaks with asymmetric Lorentzian functions over a constant background.}
\label{Tbl:J1709PlsFit}
\end{table*}

\begin{table*}[tp]
\vspace{0.2cm}
\begin{center}
\begin{tabular}{l|cccccc}
  \hline\noalign{\smallskip}
   Energy Band   & P1 Location     & P1 FWHM          & P2 Location      & P2 HWHM (Inner)    & P2 HWHM (Outer) \\
   \hline\noalign{\smallskip}
0.1 $-$ 100 GeV  & 0.154$\pm$0.001 & 0.086$\pm$0.005  &  0.639$\pm$0.004 &  0.119$\pm$0.011 & 0.034$\pm$0.007 \\
$>$3.0 GeV       & 0.156$\pm$0.008 & 0.047$\pm$0.022  &  0.650$\pm$0.004 &  0.097$\pm$0.017 & 0.006$\pm$0.002 \\
1.0 $-$ 3.0 GeV  & 0.155$\pm$0.002 & 0.034$\pm$0.005  &  0.635$\pm$0.005 &  0.122$\pm$0.012 & 0.029$\pm$0.007 \\
0.3 $-$ 1.0 GeV  & 0.155$\pm$0.002 & 0.073$\pm$0.006  &  0.640$\pm$0.004 &  0.112$\pm$0.015 & 0.035$\pm$0.010 \\
0.1 $-$ 0.3 GeV  & 0.152$\pm$0.004 & 0.127$\pm$0.016  &  0.650$\pm$0.008 &  0.126$\pm$0.027 & 0.017$\pm$0.007 \\
\hline
\end{tabular}
\end{center}
\caption{The location and Lorentzian full-width-at-half-maximum (FWHM) and half-width-at-half-maximum (HWHM) values for the first and second peaks (P1 and P2) of the \PSRThree\ pulse profiles in every energy band. These parameter values were obtained fitting the peaks with asymmetric Lorentzian functions over a constant background.}
\label{Tbl:J1952PlsFit}
\end{table*}

\begin{sidewaystable}[tp]
\begin{center}
\resizebox{22cm}{!}{
\begin{tabular}{c|cccc|cc}
\hline
PSR       & Prefactor (K)   & $\Gamma$     &  $E_{\rm Cutoff}$   & $b$   &  $F_{100}$  & $G_{100}$ \\[3pt]
             & ($10^{-10}$ ph\,cm$^{-2}$\,s$^{-1}$\,MeV$^{-1}$)
                               &              &    (GeV)          &     & ($10^{-7}$ ph\,cm$^{-2}$\,s$^{-1}$)
                                                                                           & ($10^{-4}$ MeV\,cm$^{-2}$\,s$^{-1}$)\\[3pt]
\hline\noalign{\smallskip}
J1057$-$5226  & $1.25 \pm 0.07 \pm 0.05$  & $1.20 \pm 0.05 \pm 0.11$ & $1.50 \pm 0.09 ^{+0.13}_{-0.09}$  & 1 (fixed)                & $3.36 \pm 0.10 \pm 0.20$   & $1.77 \pm 0.01 \pm 0.16$ \\[8pt]
J1057$-$5226  & $1.26 \pm 0.03 \pm 0.30$  & $1.19 \pm 0.11 \pm 0.32$ & $1.48 \pm 0.44 ^{+0.42}_{-0.22}$  & $0.99 \pm 0.15 \pm 0.07$ & 3.36$\pm$0.11$\pm$0.25     & $1.77 \pm 0.01 \pm 0.16$ \\[8pt]
J1709$-$4429  & $2.64 \pm 0.05 \pm 0.15$  & $1.71 \pm 0.02 \pm 0.06$ & $4.45 \pm 0.23 ^{+1.14}_{-0.81}$  & 1 (fixed)                & $15.97 \pm 0.24 \pm 1.12$ & $6.98 \pm 0.08 \pm 0.90$ \\[8pt]
J1709$-$4429  & $8.19 \pm 0.96 \pm 0.46$  & $1.30 \pm 0.04 \pm 0.06$ & $0.57 \pm 0.11^{+0.15}_{-0.10}$  & $0.51 \pm 0.02 \pm 0.04$  & $15.23 \pm 0.24 \pm 0.99$  & $6.99 \pm 0.09 \pm 0.71$ \\[8pt]
J1952$+$3252  & $0.39 \pm 0.03 \pm 0.02$              & $1.57 \pm 0.07 \pm 0.15$ & $2.80 \pm 0.37 ^{+0.97}_{-0.53}$  & 1 (fixed) & $1.89 \pm 0.12 \pm 0.10$  & $0.92 \pm  0.03 \pm 0.08$ \\[8pt]
J1952$+$3252  & $1.24 \pm 0.37 \pm 0.08$              & $1.13 \pm 0.31 \pm 0.12$            & $0.50 \pm 0.02 ^{+0.17}_{-0.10}$                               & $0.57 \pm 0.03 \pm 0.06$    & $1.78 \pm 0.15 \pm 0.08$  & $0.91 \pm  0.16 \pm 0.09$ \\[8pt]
\hline
\end{tabular}
}
\end{center}
\caption{Spectral parameters obtained from fitting the energy spectra with a power-law with exponential cutoff.
The first four parameters are as defined in Equation~(\ref{expcutoff}) and the last two parameters are the  integrated photon flux and the integrated energy flux above 100 MeV. The first errors are statistical and the second are systematic errors calculated from the bracketing IRFs.}
\label{Tbl:SpectResults}
\end{sidewaystable}

\begin{table*}[tp]
  \begin{center}
    \begin{tabular}{ll|lll}
      \hline\noalign{\smallskip}
      $\varphi_{\rm min}$ & $\varphi_{\rm max}$ &  Index &  $E_{\rm cutoff}$ & $F(> 100$\,MeV) $^{(a)}$                \\
                      &                 &        &   (GeV)       & ($10^{-7}$cm$^{-2}$ s$^{-1}$) \\[2pt]
      \hline\noalign{\smallskip}
0.28 & 0.32 & 1.26  $\pm$  0.11 & 1.21  $\pm$  0.19 &  13.56 $\pm$  1.13 \\
0.32 & 0.37 & 0.99  $\pm$  0.12 & 1.12  $\pm$  0.15 &  10.04 $\pm$  0.88 \\
0.37 & 0.41 & 1.17  $\pm$  0.12 & 1.42  $\pm$  0.22 &  10.61 $\pm$  0.96 \\
0.41 & 0.45 & 1.02  $\pm$  0.11 & 1.33  $\pm$  0.18 &  10.96 $\pm$  0.95 \\
0.45 & 0.49 & 0.99  $\pm$  0.12 & 1.28  $\pm$  0.19 &  11.80 $\pm$  1.05 \\
0.49 & 0.53 & 1.31  $\pm$  0.10 & 1.71  $\pm$  0.28 &  12.77 $\pm$  1.08 \\
0.53 & 0.58 & 0.91  $\pm$  0.12 & 1.29  $\pm$  0.17 &  9.01 $\pm$  0.82  \\
0.58 & 0.64 & 1.26  $\pm$  0.11 & 2.04  $\pm$  0.37 &  7.26 $\pm$  0.71  \\
      \hline
    \end{tabular}
    \tablenotetext{a}{Normalized to the full phase range.}
  \end{center}
  \caption{Phase-resolved spectral parameters for \PSROne. Only the statistical errors are quoted.}
  \label{B1055_PhaseScan_Table}
\end{table*}

\begin{table*}[tb]
\begin{center}
\begin{tabular}{cc|ccc}
\hline\noalign{\smallskip}
 $\varphi_{\rm min}$ & $\varphi_{\rm max}$ & Index  & $E_{\rm cutoff}$   & $F(>100$\,MeV) $^{(a)}$\\
                 &                 &        &   (GeV)        & ($10^{-7}$cm$^{-2}$ s$^{-1}$) \\[2pt]
 \hline\noalign{\smallskip}
0.085	& 0.165	& $2.24	\pm 0.42$ & $1.68 \pm 1.58$ &	$3.49  \pm 0.91$ \\
0.165	& 0.216	& $1.57	\pm 0.15$ & $1.14 \pm 0.23$ &	$15.25 \pm 1.21$ \\
0.216	& 0.242	& $1.73 \pm 0.07$ & $2.94 \pm 0.50$ &	$47.16 \pm 2.14$ \\
0.242	& 0.265	& $1.57	\pm 0.06$ & $3.42 \pm 0.48$ &   $52.43 \pm 2.23$ \\
0.265	& 0.292	& $1.58	\pm 0.06$ & $3.71 \pm 0.56$ &	$41.56 \pm 1.86$ \\ 
0.292	& 0.324	& $1.70	\pm 0.06$ & $5.74 \pm 1.04$ &	$36.06 \pm 1.69$ \\ 
0.324	& 0.354	& $1.63	\pm 0.06$ & $5.48 \pm 0.94$ &	$33.99 \pm 1.66$ \\
0.354	& 0.385	& $1.56	\pm 0.06$ & $3.86 \pm 0.58$ &	$33.01 \pm 1.57$ \\
0.385	& 0.416	& $1.48	\pm 0.06$ & $3.69 \pm 0.53$ &	$32.83 \pm 1.55$ \\
0.416	& 0.442	& $1.55	\pm 0.06$ & $3.87 \pm 0.57$ &	$43.89 \pm 1.92$ \\ 
0.442	& 0.466	& $1.56	\pm 0.06$ & $4.23 \pm 0.60$ &	$46.22 \pm 1.98$ \\
0.466	& 0.489	& $1.61	\pm 0.05$ & $4.74 \pm 0.71$ &	$54.43 \pm 2.22$ \\
0.489	& 0.514	& $1.65	\pm 0.05$ & $4.71 \pm 0.73$ &	$46.56 \pm 1.99$ \\ 
0.514	& 0.554	& $1.75	\pm 0.08$ & $3.18 \pm 0.67$ &	$25.90 \pm 1.49$ \\
0.554	& 0.624	& $1.72	\pm 0.23$ & $1.34 \pm 0.51$ &   $7.15  \pm 0.90$ \\
\hline
\end{tabular}
\tablenotetext{a}{Normalized to the full phase range.}
\end{center}
\caption{Phase-resolved spectral parameters for \PSRTwo. Only the statistical errors are quoted.}
\label{Tbl:SpectRsltJ1709}
\end{table*}

\begin{table*}[tb]
\begin{center}
\begin{tabular}{cc|ccc}
\hline\noalign{\smallskip}
 $\varphi_{\rm min}$ & $\varphi_{\rm max}$ & Index & $E_{\rm cutoff}$    & $F(>100$\,MeV) $^{(a)}$\\
                 &                 &        &   (GeV)        & ($10^{-7}$cm$^{-2}$ s$^{-1}$) \\[2pt]
\hline\noalign{\smallskip}
  0.094&0.155 &  1.69$\pm$0.15 & 2.34$\pm$0.76  & 5.57$\pm$2.30 \\
  0.155&0.224 &  2.63$\pm$0.15 & 1.67$\pm$0.44  & 5.65$\pm$0.58 \\
  0.490&0.559 &  1.32$\pm$0.21 & 2.80$\pm$0.93  & 2.61$\pm$0.43 \\
  0.559&0.617 &  1.57$\pm$0.11 & 3.68$\pm$0.93  & 6.72$\pm$0.52 \\
  0.617&0.673 &  1.59$\pm$0.09 & 3.84$\pm$0.84  & 7.86$\pm$1.79 \\
 \hline
\end{tabular}
\tablenotetext{a}{Normalized to the full phase range.}
\end{center}
\caption{Phase-resolved spectral parameters  for \PSRThree. Only the statistical errors are quoted.}
\label{tab:j1952interval}
\end{table*}

\begin{table*}[tb]
\vspace{0.4cm}
\begin{center}
\begin{tabular}{l|cccc}
\hline
Pulsar & $\Delta^{a}$ & $\delta^{a}$ & $\alpha(\degree)$ & $\zeta(\degree)$ \\
\hline\noalign{\smallskip}
\PSROne\   & $0.28 \pm 0.03$    & $0.31 \pm 0.03$   & $75^b$ & $69^b$ \\
\PSRTwo\   & $0.250 \pm 0.004$  & $0.242 \pm 0.002$ & -      & $53^c$ \\
\PSRThree\ & $0.485 \pm 0.017$  & $0.154 \pm 0.001$ & -      &   -    \\
\hline
\end{tabular}
\end{center}
\tablenotetext{a}{Derived from the fits on the pulsed profiles discussed in Section~\ref{Sec:PlsProfile}}
\tablenotetext{b}{Measured from radio RVM fit \citep{Weltevrede2009}}
\tablenotetext{c}{Measured from torus fits to X-ray images \citep{Ng2008}}
\caption{The measured geometric parameters for these pulsars. The parameters derived from the fits to the pulse profile, the peak separation and phase lag between the radio and the first gamma-ray pulse, are listed in the second and third columns. The $\alpha$ parameter, obtained from the radio RVM for \PSROne, is given in the fourth column and the $\zeta$ parameter, as measured from torus fits to the X-ray image for \PSRTwo\ and from the RVM fit for \PSROne, is given in the last column.}
\label{Tbl:MeasPars}
\end{table*}

\begin{table*}[tb]
\vspace{0.4cm}
\begin{center}
\resizebox{16.5cm}{!}{
\begin{tabular}{l|c|cccc|cccc}
\hline
Pulsar        & $w^{a}$
                    & $\alpha_{\rm{TPC}}$(\degree)
                              & $\zeta_{\rm{TPC}}$(\degree)
                                      & $\chi^{a}$
                                         & $f_{\Omega, \rm{TPC}}$
                                                      & $\alpha_{\rm{OG}}$(\degree)
                                                                & $\zeta_{\rm{OG}}$(\degree)
                                                                      & $\chi^{a}$
                                                                          & $f_{\Omega, \rm{OG}}$ \\[1pt]
\hline\noalign{\smallskip}
\PSROne\   & 0.183 & $75$  & $69$  & $414$  & $0.91$  & $75$  & $69$  & $106$ & $0.82$  \\
\PSRTwo\   & 0.017 & $36$  & $53$  & $155$  & $1.30$  & $36$  & $56$  & $30$  & $0.89$  \\
\PSRThree\ & 0.016 & $71$  & $84$  & $19$   & $1.10$  & $66$  & $78$  & $14$  & $0.84$  \\[1pt]
\hline
\end{tabular}
}
\end{center}
\tablenotetext{a}{$\chi$, the fit statistics as defined in \citet{Romani2009} indicates the goodness of the fits.  }
\caption{The predicted geometric parameters. The gap width $w$ from the assumption of Equation~(1) in \citet{watters09} is given in Column 2. The predicted geometric parameters,  $\alpha$ and $\zeta$ and $f_{\Omega}$, for the high altitude models, two-pole caustic (TPC) and outer gap (OG), and the goodness of fit parameter $\chi$ are from \citet{Romani2009} and they are listed in Columns $3 - 10$.  }
\label{Tbl:AtlasPars}
\end{table*}

\clearpage
\newpage

\end{document}